\documentclass[12pt,preprint]{aastex}

\shorttitle{NONTHERMAL EMISSION FROM CLUSTERS}
\shortauthors{FUJITA \& SARAZIN}
\slugcomment{NAOJ-Th-Ap 2001, No.50}

\begin{document}

\title{Nonthermal Emission from Accreting and Merging Clusters of
Galaxies}

\author{Yutaka Fujita\altaffilmark{1}}
\affil{National
Astronomical Observatory, Osawa 2-21-1, Mitaka, Tokyo 181-8588, Japan}
\email{yfujita@th.nao.ac.jp}

\and

\author{Craig L. Sarazin}
\affil{Department of Astronomy, University of Virginia, P.O. Box 3818, 
Charlottesville, VA 22903-0818} 

\email{sarazin@virginia.edu}

\altaffiltext{1}{Present address: Department of Astronomy, University of 
Virginia, P.O. Box 3818, Charlottesville, VA 22903-0818}

\begin{abstract}
We compare the nonthermal emission from clusters of galaxies undergoing
minor mergers (``accreting'' clusters) and major mergers (``merging''
clusters).  We define major mergers as the mergers that change the inner
dark matter structure of clusters; minor mergers are all others. For
accreting clusters, the radial distribution of the nonthermal emission
in the clusters is also calculated.  The relativistic electrons, which
are the origin of the nonthermal radiation through inverse Compton (IC)
and synchrotron emission, are assumed to be accelerated at shocks
produced by accretion or mergers. We estimate the typical accretion rate
and merger probability according to a hierarchical clustering model.  We
predict that in the inner region of accreting clusters the nonthermal
emission has a flat spatial distribution at all frequency.  For
synchrotron and hard X-ray emissions, we predict an increase of the
emissions at the cluster edge due to accretion.  We show that the total
luminosity of IC emission from accreting and merging clusters are
similar.  On the other hand, the luminosity of synchrotron radio
emission of the former is much smaller than that of the latter.  We show
that about 10\% of clusters at $z\sim 0$ should have hard X-ray and
radio nonthermal emissions due to their last major merger, which are
comparable to or dominate those due to ongoing accretion. Moreover,
$20-40$\% of clusters should have significant EUV emission due to their
last merger.  We also investigate the case where the criterion of
mergers is relaxed. If we extend the definition of a merger to an
increase in the mass of the larger subcluster by at least 10\% of its
initial mass, about $20-30$\% of clusters at $z\sim 0$ should have hard
X-ray and radio nonthermal emissions due to the merger even in a low
density universe.  We compare the results with observations. We find
that the observed EUV emission from clusters is not attributed to
accretion. If the diffuse radio emission observed in clusters is
synchrotron emission from electrons accelerated via accretion or
merging, the magnetic fields of clusters are generally as small as $\sim
0.1\rm\: \mu G$. One concern is that this is a significantly weaker
field than that implied by Faraday rotation measurements.
\end{abstract}


\section{Introduction}

Clusters of galaxies are luminous, extended X-ray sources.
The bulk of this radiation is thermal emission (mainly thermal
bremsstrahlung and line emission) from hot gas with temperatures of
$\sim$7 keV (the intracluster medium or ICM).
On the other hand, recent observations at very soft X-ray/extreme
ultraviolet (EUV) energies ($\sim$0.1 keV) and at very hard X-ray
energies ($\sim$20 keV) suggest excess emission beyond that expected
 from the thermal ICM emission.

Extreme ultraviolet (EUV) or very soft X-ray excess emissions have been
detected from a number of clusters
\citep{lie96a,lie96b,bow96,bow98,mit98,kaa99,lie99a,lie99b,ber00a,
bon01}. Although the properties and even the existence of this EUV
emission is still controversial \citep{ara99,bow99,ber00b}, one
hypothesis is that it is the inverse Compton (IC) scattering of cosmic
microwave background (CMB) photons by relativistic electrons
\citep{hwa97,bow98,ens98a,sar98}.  These electrons would have energies
of $E = \gamma m_c c^2 \sim 150$ MeV and Lorentz factors $\gamma \sim
300$.

Hard X-ray emission in excess of the thermal emission and detected as a
nonthermal tail at energies $\ga$20 keV has been seen in at least two
clusters.  The Coma cluster was detected with both {\it BeppoSAX}
\citep{fus99} and the {\it Rossi X-Ray Timing Explorer} ({\it RXTE} ;
\citealp*{rep99}). {\it BeppoSAX} also has detected Abell~2256
\citep{fus00}.  Possible weaker excesses may have been seen in
Abell~2199 \citep{kaa99} and Abell~3667 \citep{fus01}.  Again, this
radiation is believed to be IC emission, in this case produced by
relativistic electrons with $E = \gamma m_c c^2 \sim 5$ GeV and Lorentz
factors $\gamma \sim 10^4$.

A number of clusters are known to have diffuse radio sources
\citep[e.g.,][]{kim90,gio93,gio00,kem01}.  These radio sources have very
steeply declining radio spectra, and are not associated with individual
galaxies.  They are referred to as radio halos when they appear
projected on the center of the cluster, and are called relics when they
are found on the cluster periphery (although they have other distinctive
properties).  The origin of the radio emission is definitely nonthermal,
being synchrotron emission produced by relativistic electrons radiating
in the intracluster magnetic fields.  The typical energy of the
electrons which produce the radio emission depends on the observed
frequency and on the intracluster field, but is estimated to be several
GeV; thus, basically the same electrons which produce the IC hard X-ray
emission produce this diffuse radio emission

The relativistic electrons are often believed to be accelerated by
shocks in ICM \citep{jaf77,rol81,sch87}, although alternative origins
are possible \citep{ato00,ens01}.  These shocks may be attributed to the
interaction between jets originated from active galactic nuclei (AGNs)
and ICM.  On the other hand, given the large size of the the diffuse
radio relics and halos and the extent of EUV/soft X-ray emission in
clusters, it may be easier to understand these sources if the particles
are produced by mergers between clusters \citep{roe99}. \citet{tak00}
calculated the nonthermal emission from relativistic electrons
accelerated around the shocks produced during a merger of clusters with
equal mass. They found that the hard X-ray and radio emissions are
luminous only while signatures of merging events are clearly seen in the
ICM. On the other hand, EUV emission is still luminous after the system
has relaxed.

In this paper, we will divide cluster mergers into two groups: mergers
that lead to a mass increase by as much as the initial mass of main
subcluster and to the reconstruction of cluster dark matter structure,
and the mergers of smaller subclusters into a much larger main cluster.
We refer to the former process as a ``merger,'' and the latter as
``accretion''.  \citet{kau93} indicated that the percentage of clusters
undergoing merging events with roughly equal-mass progenitors over a
1~Gyr period at low redshift is 3-15 percent. Thus the results of
\citet{tak00} can be compared with the observations of only a small
fraction of clusters. On the other hand, \citet{kau93} showed that most
clusters are accreting smaller clumps at present.  Through this
accretion, shocks form around the virial radius of clusters
\citep{evr96,tak98,eke98,min00}. \citet{kan97} modeled the acceleration
of very high energy particles around the shock at the cluster edge.
Nonthermal emission may be produced by the accelerated electrons at the
cluster accretion shock, although the number of such electrons and the
emission power may be small compared with the case of violent mergers.
In the future, it may be possible to observe fainter nonthermal emission
 from clusters, and to study the nonthermal radiation from large samples
of clusters statistically.  Thus, it would be important to study the
nonthermal emission not only from rare clusters undergoing violent
merging events but also that from normal clusters undergoing accretion.

In this paper, we study the nonthermal emission from accreting clusters
and compare the results with the emission from clusters undergoing major
mergers. \citet{sar99} considered a simple model for the time evolution
of the relativistic electrons and nonthermal emission from clusters
accreting ambient medium, based on the self-similar accretion solution
of \citet{ber85}.  However, this paper did not directly compare the
expected fluxes of merging and accreting clusters, as we do here.
\citet{ens98b} proposed that the shocks caused by accretion onto
clusters and the resulting acceleration of electrons in the shocks are
responsible for observed cluster radio relics using the results of one
dimensional simulations of cluster formation in the Einstein-de Sitter
universe. However, they mainly focused on the emission of individual
radio relics and did not consider EUV and hard X-ray emissions. We
consider EUV, hard X-ray and radio emissions from clusters undergoing
accretion or merger generally. The somewhat arbitrary distinction
between accretion and major mergers may be useful, because they differ
considerably in the the power and spatial distribution of the resultant
nonthermal emission.  For example, while there is evidence that the
strongest radio halos appear only in those clusters currently
experiencing major mergers \citep[e.g.,][]{buo01}, many other clusters
may have weaker diffuse radio emission, below the sensitivity of current
surveys, which is associated with weaker mergers (accretion).  This
paper is organized as follows: \S~\ref{sec:model}, we summarize our
models.  We give the results of our calculations in \S~\ref{sec:res},
and compare them with observations in \S~\ref{sec:dis}.  Conclusions are
given in \S~\ref{sec:conc}.

\section{Models}
\label{sec:model}
\subsection{Mass Evolution of a Cluster}

In order to follow the evolution of clusters or, more generally, `dark
halos,' we use a modification of an extended Press \& Schechter model
derived by \citet{sal98}. From now on, we call the model the SSM model.
In this model, major mergers among clusters and continuous accretion are
distinguished. Here, major mergers mean the mergers that change the core
structure of dark halos; minor mergers are all others. We will refer to
major mergers as `mergers', and will use the term `accretion' to
describe the cumulative effect of minor mergers. 
Note that a `merger' does not necessary mean a binary merger; it also
includes multiple mergers. The details of the model are also described
in \citet{eno01}.

In this model, the mass function of dark halos is given by the Press \&
Schechter (PS) mass function;
\begin{equation}
n(M,t) \, dM = \sqrt{\frac{2}{\pi}} \frac{\rho_0}{M} 
\frac{\delta_c(t)}{\sigma^{2}(M)}
\left|  \frac{d\sigma(M)}{dM} \right| 
\exp \left[-\frac{1}{2} \frac{\delta_c^{2}(t)}{\sigma^{2}(M)} \right] \, dM
\, ,
\label{eq:PS}
\end{equation}
where $n(M,t) \, dM$ gives the number density of clusters with masses in
the range $M \rightarrow M + d M$ per comoving volume.  Here, $\rho_0$
is the present mean density of the universe, $\delta_c(t)$ is the
critical density contrast for collapse at $t$, and $\sigma(M)$ is the
rms density fluctuation in spheres containing a mean mass $M$
\citep{pre74}. In this paper, we use an approximate formula of
$\delta_c(t)$ for spatially flat cosmological model \citep{nak97} and a
fitting formula of $\sigma(M)$ for the CDM fluctuation spectrum
\citep{kit97}.

\citet{lac93} obtained the instantaneous merger rate for halos with mass
$M$ at time $t$ per infinitesimal range of final mass $M' > M$; it is
given by
\begin{eqnarray}
r_{\rm LC}^{m}(M \to M',t) \, dM' & \equiv & \sqrt{\frac{2}{\pi}} \ \left|
\frac{d\delta_c(t)}{dt} \right| \ \frac{1}{\sigma^{2}(M')} \ \left|
\frac{d\sigma(M')}{dM'} \right| \nonumber \\ & \times & \left[1-
\frac{\sigma^{2}(M')}{\sigma^{2}(M)} \right]^{- 3/2} \nonumber \\ &
\times & \exp \left\{ -\frac{\delta_c^{2}(t)}{2} \left
[\frac{1}{\sigma^{2}(M')}-\frac{1}{\sigma^{2}(M)} \right] \right\} \, dM'
\, .
\label{eq:LCmerge}
\end{eqnarray}
In the SSM model, it is assumed that a halo with mass $M$ experiences a
merger and is destroyed when the relative mass increment $\Delta M /M
\equiv (M'-M)/M$ exceeds a certain threshold $\Delta_m$.  The merger is
regarded as the formation of a new halo.  On the other hand, when
$\Delta M /M <\Delta_m$, the event is regarded as continuous accretion;
the halo keeps its identity and its structure. Thus, from the specific
merger rate (equation [\ref{eq:LCmerge}]), the mass accretion rate,
$R_{\rm mass}(M,t) \equiv dM/dt$, of halos with mass $M$ at time $t$ is
defined as
\begin{equation}
R_{\rm mass}(M,t) 
= \int_{M}^{M(1+\Delta_m)} \Delta M \, r_{\rm LC}^{m}(M \to M'
 ,t) \, dM' \, .
\label{eq:R-mass}
\end{equation}
The destruction rate of halos is defined as
\begin{equation}
r^{d}(M,t) = \int_{M(1+\Delta_m)}^{\infty} r_{\rm LC}^{m}(M \to M',t) \, dM'
\, .
\label{eq:destrate}
\end{equation}
On the other hand, the formation rate of halos is given by
\begin{equation}
r^{f}[M(t),t] = \frac{d \ln n[M(t),t]}{d t} + r^{d}[M(t),t] 
+ \frac{\partial R_{\rm mass}[M(t),t]}{\partial M} \, ,
\label{eq:continue}
\end{equation}
which follows from the conservation equation for the number density of
halos per unit mass along mean mass accretion tracks, $M(t)$, which is
obtained by solving the differential equation
\begin{equation}
\label{eq:M-track}
\frac{dM}{dt} = R_{\rm mass}[M(t),t] .
\end{equation}
 From the formation rate, we can obtain the distribution of formation
times for halos with masses $M_0$ at the present time $t_0$;
\begin{equation}
\Phi_f(t;M_0,t_0) \, dt
= r^{f}[M(t),t] 
\exp \left\{- \int_{t}^{t_0} r_f[M(t'),t'] \, d t' \right\} dt \, ,
\label{eq:distSSM}
\end{equation}
where $\Phi_f(t;M_0,t_0) \, dt$ gives the probability that a clusters
with a mass $M_0$ at the present time ($t_0$) had its last major merger
during the time period $t \rightarrow t + dt$.  The median of this
distribution is adopted as the typical halo formation time or the last
merger time, $t_f$.

The value of $\Delta_m$ is fixed by the fits to the empirical
mass-density (or mass-radius) correlation obtained by $N$-body
simulations \citep{nav97}. \citet{sal98} showed that the best fit is
$\Delta_m = 0.6$ in a number of different cosmological models.  In other
words, the core structure of halos are destroyed when $\Delta_m \gtrsim
0.6$.  We use this value in \S\ref{sec:res}.  Although merger and
accretion cannot be distinguished clearly and the value has a range
($0.5\lesssim \Delta_m \lesssim 0.7$), the simple distinction would make
the results of calculations easy to understand.  On the other hand,
hydrodynamical simulations of clusters show that smaller mergers can
still produce internal shocks and violent hydrodynamical activity in
clusters especially when they are binary mergers (e.g., the $\Delta M /
M = 1/8$ mergers in \citeauthor{roe99} [1999] or the $\Delta M / M =
1/3$ mergers in \citeauthor{ric01} [2001]).  These smaller mergers
($0.1\lesssim \Delta M / M \lesssim 0.6$) are often observed as
``mergers'', although they do not significantly change the dark matter
structure of main clusters {\it in the end}; they only extend the
density distribution \citep{sal98,nav97}. Since ICM is in pressure
equilibrium with the dark matter potential except during the mergers, we
expect that those ``semi-mergers'' do not significantly change the ICM
structure in the end, either.  In \S\ref{sec:res}, we will discuss the
nonthermal emission from the semi-mergers, which are included in
accretion in our classification unless otherwise mentioned.

\subsection{Shock Acceleration of Electrons at the Cluster Edge}

The virial radius of a cluster with virial mass $M(t)$ is defined
as
\begin{equation}
 \label{eq:r_vir}
r_{\rm vir}(t)=\left[\frac{3\: M(t)}
{4\pi \Delta_c(t) \rho_{\rm crit}(t)}\right]^{1/3}\:,
\end{equation}
where $\rho_{\rm crit}(t)$ is the critical density of the universe and
$\Delta_c(t)$ is the ratio of the average density of the cluster to the
critical density at time $t$. The former is given by
\begin{equation}
\label{eq:rho_crit}
 \rho_{\rm crit}(t)
=\frac{\rho_{\rm crit,0}\Omega_0 (1+z)^3}{\Omega(t)}\:,
\end{equation} 
where $\rho_{\rm crit,0}$ and $\Omega_0 = \rho_0 / \rho_{\rm crit,0}$
are the critical density and cosmological density parameter at present,
and $\Omega(t)$ is the cosmological density parameter, and $z$ is the
redshift corresponding to time $t$. The latter is given by
\begin{equation}
\label{eq:Dc_lam}
  \Delta_c(t)=18\:\pi^2+82 x-39 x^2\:, 
\end{equation}
for a flat universe with non-zero cosmological constant
\citep{bry98}. In equation (\ref{eq:Dc_lam}), the parameter $x$ is given
by $x=\Omega(t)-1$.

We assume that a cluster is spherically symmetric and that a shock is
formed at the virial radius of a cluster \citep{evr96,tak98,eke98}.
Since we adapt a rather large upper mass cutoff for accretion, and
hydrodynamical simulations show that shocks from smaller mergers
penetrate much deeper into the cluster interior than $r_{\rm vir}$
(e.g., the $\Delta M / M = 1/8$ mergers in \citeauthor{roe99} [1999] or
the $\Delta M / M = 1/3$ mergers in \citeauthor{ric01} [2001]), this
assumption overestimates the radius at which nonthermal effects due to
accretion occur.  However, as will be shown in \S\ref{sec:acc}, the
contribution of such semi-mergers to the nonthermal emission due to
accretion is not large in general. For continuous accretion, spherical
models indicate that the structure of a cluster does not change much
inside the shock radius \citep{tak98}.  Thus, we assume that the mass of
a cluster between radii $r=r_{\rm vir}(t_i)$ and $r_{\rm vir}(t_i+\delta
t_i)$ is $M(t_i+\delta t_i)-M(t_i)$ for $t_i>t_f$. The virial mass of a
cluster is obtained by solving equation (\ref{eq:M-track}) for a given
virial mass at $z=0$. The gravitational energy imparted to the baryon
contained in the mass shell is approximately given by
\begin{equation}
\label{eq:Eb}
E_{\rm b}[r_{\rm vir}(t_i),r_{\rm vir}(t_i+\delta t_i)]
\approx f_b\frac{1}{2}v_{\rm vir}(t_i)^2 [M(t_i+\delta
t_i)-M(t_i)] \:,
\end{equation}
where $f_b$ is the baryon fraction.  We use $f_{\rm b}=0.25
(h/0.5)^{-3/2}$, where the present value of the Hubble constant is
written as $H_0=100\:h\rm\: km\:s^{-1}\: Mpc^{-1}$. The value of $f_b$
is the observed ICM mass fraction of high-temperature clusters
\citep{moh99,ett99,arn99}, for which the effect of heating other than
gravity is expected to be small \citep[e.g.,][]{cav98,fuj00,loe00}. In
equation (\ref{eq:Eb}), $v_{\rm vir}$ is the velocity of the mass shell
just before it reaches $r=r_{\rm vir}(t_i)$, which is given by
\begin{equation}
 \frac{1}{2}v_{\rm vir}(t_i)^2=\frac{GM(t_i)}{r_{\rm vir}(t_i)}
-\frac{GM(t_i)}{r_{\rm ta}[r_{\rm vir}(t_i)]}\:.
\end{equation}
In this equation, $r_{\rm ta}[r_{\rm vir}(t_i)]$ is the turn around
radius of the mass shell and is given by $r_{\rm ta}[r_{\rm
vir}(t_i)]=2r_{\rm vir}(t_i)$ on the basis of virial theorem. 

We expect that some particles are accelerated to relativistic energies
by the shock formed by accretion, and that the high-energy electrons
radiate nonthermal emission. In this paper, we consider only the
electrons directly accelerated at the shock (primary electrons), and
ignore the electrons produced as secondaries by cosmic ray interactions
(secondary electrons).  This is because nonthermal protons, which are
the source of the secondary electrons, accumulate in a cluster
throughout its life \citep{bla01}; in order to calculate the evolution
of the proton population, we may need to follow the entire merging
history of the cluster.

We assume that an energy $\xi_e E_b$ goes into the accelerating
electrons. From observations of supernova remnants, we assume
$\xi_e=0.05$ \citep{koy95,tan98}.
Let $N( \gamma , t ) \, d \gamma$ be the number of relativistic
electrons with Lorentz factors in the range $\gamma \rightarrow \gamma +
d \gamma$ per unit volume, and let $N_i ( \gamma_i , t_i )$ be the
initial electron density injected by acceleration at the accretion
shock.  We assume that the particles are accelerated with a power-law
energy distribution, $N_i (\gamma_i,t_i) = N_1 \gamma_i^{-p_0}$.  The
normalization $N_1$ is given by the relation
\begin{equation}
 \int_{\gamma_{\rm min}}^{\infty} m_e c^2 \,
(\gamma_i-1) \, N_i (\gamma_i,t_i) \, d\gamma_i
=\xi_e \,
\frac{E_b[r_{\rm vir}(t_i),r_{\rm vir}(t_i+\delta t_i)]}
{ (4 \pi / 3 ) \, [r_{\rm vir}(t_i+\delta t_i)^3 - r_{\rm vir}(t_i)^3]}
\, ,
\end{equation}
where $m_e$ is the electron mass and $c$ is the light speed.
Following \citet{sar99}, we use $\gamma_{\rm min}=300$ and
$p_0=2.3$. 
Note that the particle spectrum extends to $\gamma < 300$.

Electrons accelerated at $r=r_{\rm vir}(t_i)$ and $t=t_i$ lose their
energy through inverse Compton (IC) scattering of cosmic microwave
background (CMB), synchrotron radiation, Coulomb loss, and
bremsstrahlung radiation. The energy loss rates are respectively given
by
\begin{equation}
\label{eq:IC}
 b_{\rm IC}(\gamma)=1.37\times 10^{-20}\gamma^2 (1+z)^4\rm\: s^{-1}\:,
\end{equation}
\begin{equation}
\label{eq:syn}
 b_{\rm syn}(\gamma)=1.30\times 10^{-21}
\gamma^2 \left(\frac{B}{1\rm\; \mu G}\right)^2 \rm\; s^{-1}\;,
\end{equation}
\begin{equation}
\label{eq:Coul}
 b_{\rm Coul}(\gamma)\approx 1.2\times 10^{-12}n_e 
\left[1.0+\frac{\ln (\gamma/n_e)}{75}\right]\rm\; s^{-1}\:,
\end{equation}
\begin{equation}
\label{eq:brems}
 b_{\rm brem}(\gamma)\approx 1.51\times 10^{-16}n_e 
\gamma[\ln (\gamma)+0.36] \rm\: s^{-1}\:,
\end{equation}
where $B$ is the magnetic field and $n_e$ is the thermal electron
density \citep{blu70,rep79,sar99}.  The evolution of the energy of an
electron is described by
\begin{equation}
 \frac{d\gamma}{dt}=-b(\gamma, t)\:,
\end{equation}
where $b=b_{\rm IC}+b_{\rm syn}+b_{\rm Coul}+b_{\rm brem}$. We assume
that the accelerated electrons remain in the same place. (In
\S\ref{sec:acc}, we discuss the validity of this assumption.) Moreover,
we assume that there is no additional injection of particles at a
certain radius after a shock moves outwards. Thus, the differential
population density in a mass shell for $t>t_i$ is given by
\begin{equation}
 N(\gamma,t)=N_i(\gamma_i,t_i)\left.\frac{\partial \gamma_i}{\partial 
\gamma}\right|_t\;.
\end{equation}

The electron density in a mass shell between $r_{\rm vir}(t_i)$ and
$r_{\rm vir}(t_i+\delta t_i)$ is assumed to be
\begin{equation}
\label{eq:ne}
 n_e=\frac{3}{4\pi}\frac{0.86}{m_H}\frac{M(t_i+\delta t_i)-M(t_i)}
{r_{\rm vir}(t_i+t\delta_i)^3-r_{\rm vir}(t_i)^3}\:,
\end{equation}
where $m_H$ is the mass of hydrogen. Here, we assumed that the metal
abundance of the ICM is 0.3 solar abundance.  We assume that magnetic
field in a cluster is adiabatically compressed. Thus, the magnetic field
in the mass shell is given by
\begin{equation}
\label{eq:mag}
B = B_0 (n_e/n_{e0})^{2/3}
\end{equation}
where $n_{e0}$ is the electron density of the background universe and
$B_0$ is a parameter.  We use $B_0=0.01\rm\: \mu G$ unless otherwise
mentioned; the value is consistent with observed synchrotron emission
 from clusters (\S\ref{sec:dis}).  We assume that the magnetic field at a
fixed location does not change after the passage of the accretion
shock. Note that the cooling time of most of the accelerated electrons
is relatively short except for lower energy electrons which make EUV
emission \citep{sar99}. Thus, a slow change of the internal structure of
clusters does not strongly affect the present nonthermal emission from
the electrons except for the EUV.

\subsection{Mergers}
\label{sec:mer}

In the SSM model, a cluster is assumed to experience a merger when
$\Delta M/M>\Delta_m$.  In violent mergers, shocks will propagate
through the ICM of the merger remnant, and these shocks should
accelerate particles throughout the whole cluster. In this case, the
emission from the nonthermal particles comes not only from the outer
region but also from the central region of the cluster \citep[see
Figs.~3 and~4 in][]{tak00}. In this subsection, we construct a model to
estimate the nonthermal emission of merging clusters using a spherical
collapse model.

When a cluster is approximated by an isothermal sphere, the gravitational
energy is given by
\begin{equation}
 E_G\approx \frac{3}{4}\frac{GM^2}{r_{\rm vir}}\:.
\end{equation} 
Thus, a similar amount of energy should be released at the merger that
makes a cluster with the mass of $M$ \citep{tot00}. We consider only
the last merger, which occurred at $t=t_{f}$.  The cluster mass just
after the merger is given by $M(t_{f})$, where $M(t)$ is the solution of
equation (\ref{eq:M-track}) for a given present mass. The virial radius
$r_{\rm vir}$ is given by equation (\ref{eq:r_vir}). We assume that the
energy $E_{\rm e,mer}=\xi_e f_{\rm b}E_G$ goes to accelerated electrons
by the merger.

We assume that the shock velocity in a merging cluster is
$V_s=(4/3)V_f$, which is the velocity of a strong shock when a material
is shocked by a supersonic piston with a velocity of $V_f$.  Note that
if the Mach number of the shock is $\sim 2$, the shock velocity
increases by $\sim 30$\%. We define the velocity as $V_f\approx
\sqrt{GM/r_{\rm vir}}$ e.g., the infall velocity of material from $r= 2
r_{\rm vir}$. The time for the shock wave to propagate across the radius
of the cluster is $t_{\rm shock}=r_{\rm vir}/V_s$. Thus, during a
merger, the energy injection rate to accelerated electrons is assumed to
be $E_{\rm e,mer}/t_{\rm shock}$. The accelerated electrons lose their
energy by the same processes as in the accretion model
(equations~[\ref{eq:IC}]-[\ref{eq:brems}]).
As the electron density around the shock front in a merging cluster, we
use a typical density;
\begin{equation}
 n_e=f_b\frac{0.86 M}{m_H (4\pi/3)r_{\rm vir}^3}\;,
\end{equation}
For the magnetic field around the shock front, we use equation
(\ref{eq:mag}). We use $\xi_e=0.05$ and $B_0=0.01\rm\: \mu G$ as is the
case of the accretion unless otherwise mentioned.

\section{Results}
\label{sec:res}

\begin{figure}\epsscale{0.40}
\plotone{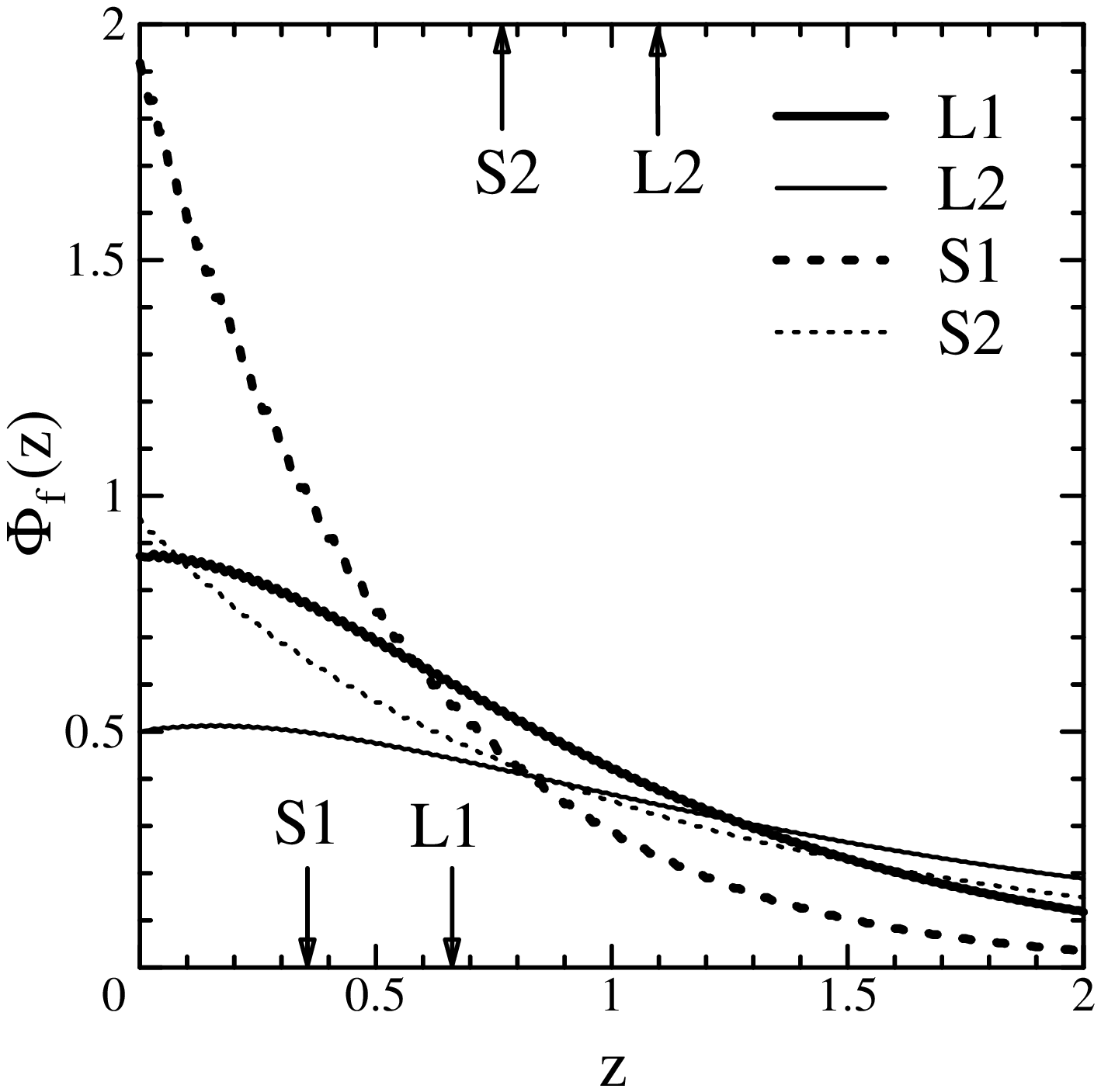} \caption{Probability distribution function $\Phi_f(z)$
for the redshift $z$ of the last merger for models~L1 (thick solid
line), L2 (thin solid line), S1 (thick dotted line), and S2 (thin dotted
line). $\Phi_f(z)$ is defined such that $\Phi_f(z)\,dz$ gives the
probability that the last major merger occurred between redshifts $z$
and $z + dz$.  The arrows indicate the median last merger redshift
$z_f=z(t_f)$. \label{fig:phi}}
\end{figure}

We consider several models with differing cosmological parameters and
values for the present mass of the cluster. Table~\ref{tab:par} shows
the density parameter $\Omega_0$, the cosmological constant $\lambda$,
the Hubble constant parameter $h$, the present rms density fluctuation
in $8\; h^{-1}\;\rm Mpc$ spheres $\sigma_8$, the present mass of a
cluster $M_0\equiv M(z=0)$, and the magnetic field $B_0$. The masses
$M_0=10^{15}\; M_{\sun}$ and $M_0=10^{14}\; M_{\sun}$ correspond to
clusters of galaxies and groups of galaxies,
respectively. Figure~\ref{fig:phi} shows the probability distributions
of the last merger of clusters. The arrows in the figure indicate the
median of the distributions or the typical redshifts at which last
mergers occurred, $z_f=z(t_f)$.

\subsection{The Emission from a Cluster Undergoing Accretion}
\label{sec:acc}

\begin{figure}\epsscale{0.40}
\plotone{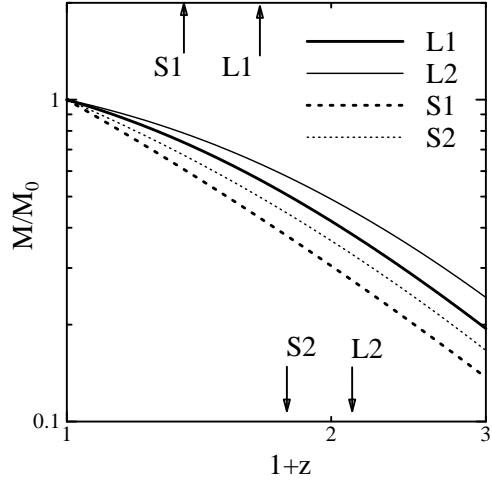} \caption{Evolution of the virial mass of a cluster for
models~L1 (thick solid line), L2 (thin solid line), S1 (thick dotted
line), and S2 (thin dotted line), assuming only accretion occurs
(eq.~\protect\ref{eq:M-track}).  The mass is normalized by the
present-day mass $M_0$. The arrows indicate the median last merger
redshift $z_f=z(t_f)$. \label{fig:m}}
\end{figure}

We solve the equation~(\ref{eq:M-track}) and derive the evolution of the
virial mass of a cluster after the last merger (Figure~\ref{fig:m}). To
emphasize accretion, the curves are computed assuming the last merger
occurred at $ z > 2$. The mass of a cluster in the Einstein-de Sitter
universe (models~S1 and S2) decreases more rapidly with redshift than in
a low density universe (models~L1 and L2), although the difference is
not large.

\begin{figure}\epsscale{0.50}
\plotone{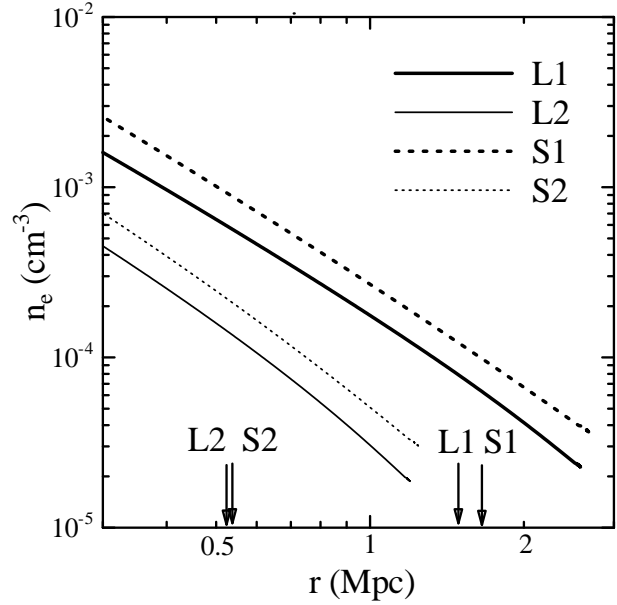} \caption{Gas density profiles for models~L1 (thick
solid line), L2 (thin solid line), S1 (thick dotted line), and S2 (thin
dotted line). The arrows indicate the virial radius at the last merger
$r_{\rm vir}(t_f)$. \label{fig:nr}}
\end{figure}

\begin{figure}\epsscale{0.50}
\plotone{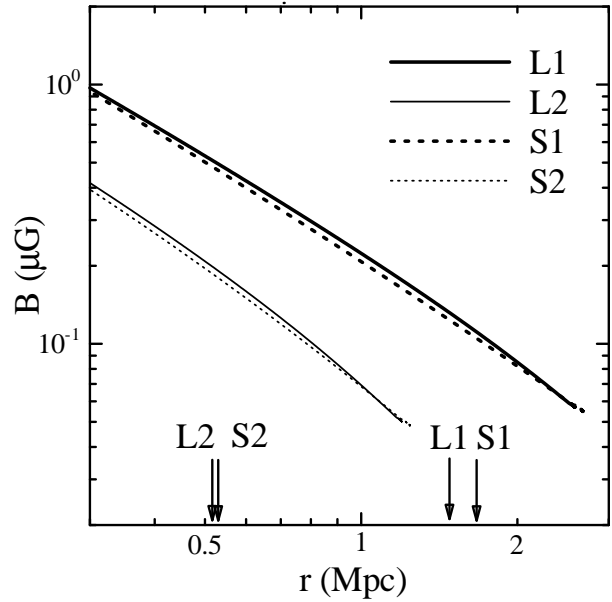} \caption{The same as Figure~\ref{fig:nr} but for
magnetic field profiles.  \label{fig:Br}}
\end{figure}

Figures~\ref{fig:nr} and~\ref{fig:Br} respectively show the
distributions of gas density and magnetic fields. The right ends of the
curves correspond to the present virial radius of clusters. The arrows
in the figures indicate $r_{\rm vir}(t_f)$.  Given the simplifying
assumptions, the density profiles may not be very realistic.  In
particular, for $r\lesssim r_{\rm vir}(t_f)$, mergers may affect the
structure of the gas.  However, since the cooling time of high energy
electrons is smaller than the time scale for changes to the cluster
structure, the inaccuracy of the density profile in the inner region of
a cluster does not affect the nonthermal emission from the
electrons. (EUV emission may be affected because electrons responsible
for it have relatively large cooling time [\citeauthor{sar99} 1999].)
Moreover, for $r\gtrsim r_{\rm vir}(t_f)$, the density profiles are
roughly $n_e\propto r^{-2.3}$, which is roughly consistent with the
outer part of the so-called universal density profile
\citep{nav96,nav97}. Since the universal density profile is often
claimed to represent the actual density profiles of dark matter well,
the obtained profiles in Figure~\ref{fig:nr} may not be unrealistic at
least for their outer parts. Note that we implicitly assumed that the
density profile of ICM follows that of dark matter or assumed that the
effect of non-gravitational heating is small. This assumption is good at
least for massive clusters \citep[e.g.,][]{cav98,fuj00,loe00}; for
smaller clusters some modification may be required but it is beyond our
scope in this paper. Since the following results are not much different
between model series~L and S, we present the results of models~L1 and L2
in the rest of this subsection.

\begin{figure}\epsscale{0.50}
\plotone{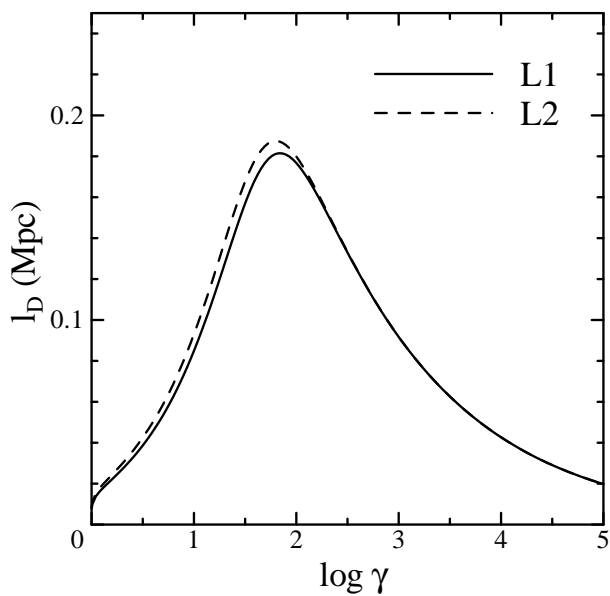} \caption{Diffusion length of an electron at the virial
radius at $z = 0$ for models~L1 (solid line) and L2 (dashed
line).\label{fig:diff}}
\end{figure}

Using the profiles of magnetic fields, we can estimate the diffusion
length of electrons. The diffusion coefficient of relativistic electrons
is given by
\begin{equation}
 D_{\rm CR}(E)\approx 5.0\times 10^{29}
\left(\frac{E}{1\:\rm GeV}\right)^{1/3}
\left(\frac{B}{0.1\:\rm \mu G}\right)^{-1/3}\rm\: cm^2\: s^{-1}\:,
\end{equation}
where $E$ is the energy of an electron \citep{ber97,col98}. The distance
which an electron reaches before being affected by cooling is given by
\begin{equation}
 l_D\approx \sqrt{6 D_{\rm DR} t_{\rm cool}}\approx
100 \left(\frac{E}{1\:\rm GeV}\right)^{1/6}
\left(\frac{B}{0.1\:\rm \mu G}\right)^{-1/6}
\left(\frac{t_{\rm cool}}{1\rm\: Gyr}\right)^{1/2}
\rm\: kpc\:,
\end{equation}
where $t_{\rm cool}$ is the cooling time of an electron. Assuming
$t_{\rm cool}=1/b(\gamma,t)$, we calculate $l_D$ and present it in
Figure~\ref{fig:diff} for an electron at $r\sim r_{\rm vir}(z=0)$ for
models~L1 and L2. Since $l_D$ is much smaller than the virial radius of
clusters, diffusion does not affect the radial profile of nonthermal
emission from the high-energy electrons significantly.  In particular,
synchrotron and hard X-ray emissions are not affected by the diffusion
because higher energy electrons ($\gamma\gtrsim 10^4$) are responsible
for them.

\begin{figure}\epsscale{0.50}
\plotone{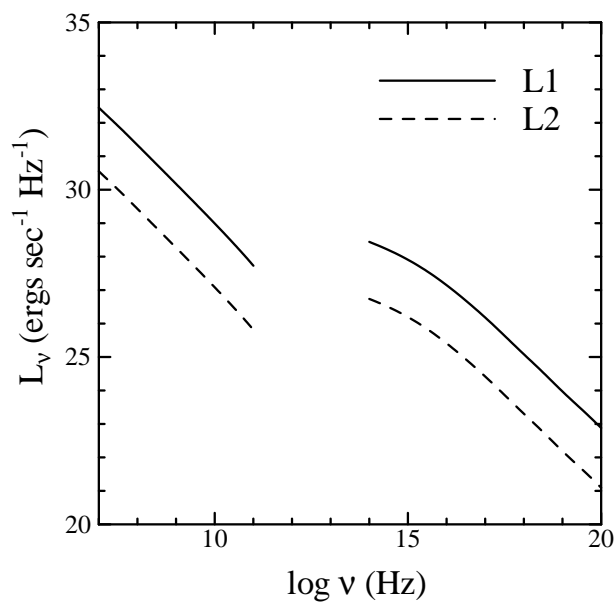} \caption{Spectra of total emission of a cluster for
models~L1 (solid line) and L2 (dashed line). The emission at
$\nu<10^{11}$~Hz is synchrotron emission and the emission at
$\nu>10^{14}$~Hz is IC emission. \label{fig:ts}}
\end{figure}

Figure~\ref{fig:ts} shows the spectra of the total emission of a
cluster.  The overall profiles are similar to those in steady injection
models in \citet{sar99}. The absolute values of the power are smaller
than those of models~1 and 20 in \citet{sar99} because of smaller
injection rate of high-energy electrons in our models. 

\begin{figure}\epsscale{1.00}
\plottwo{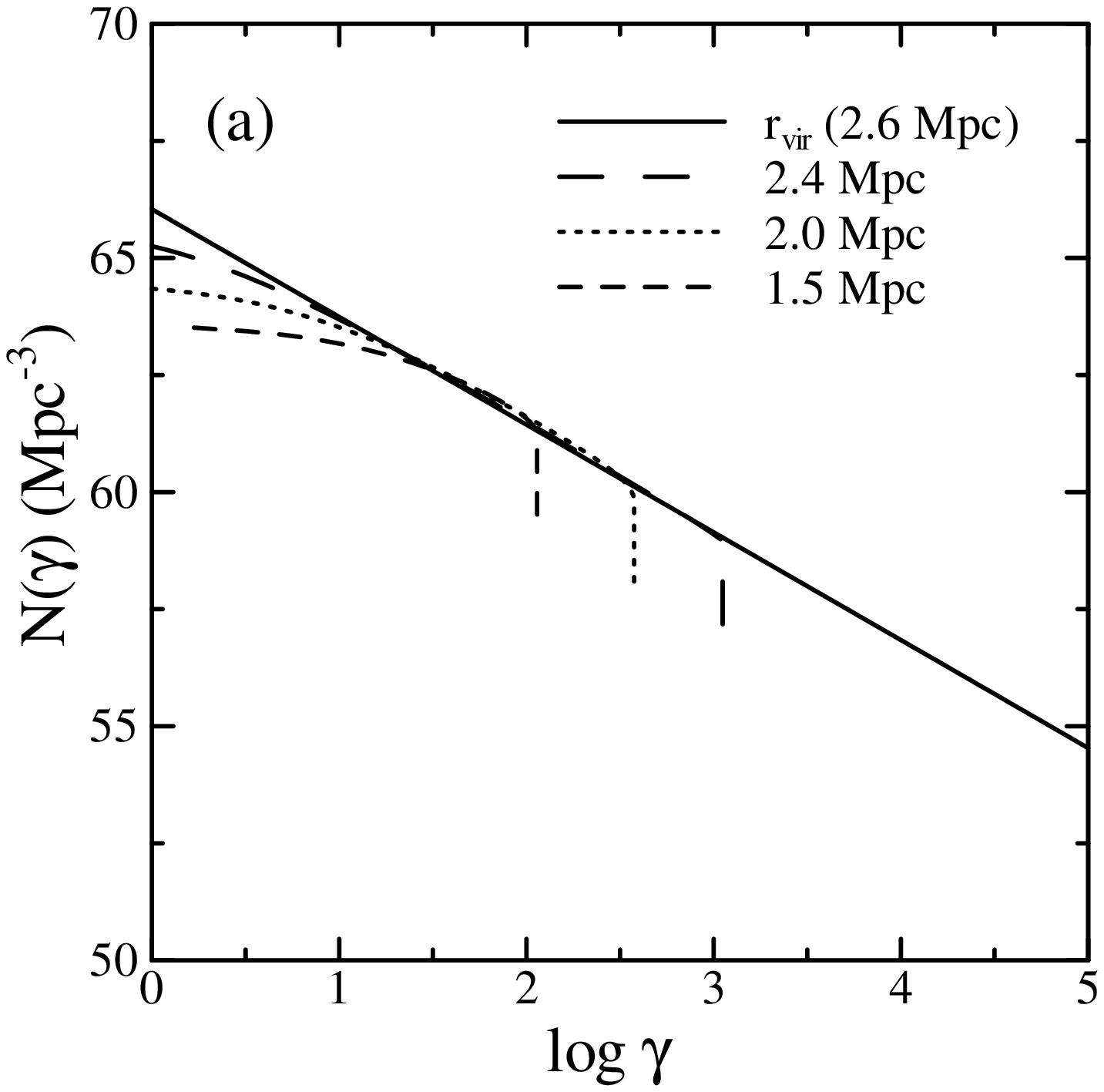}{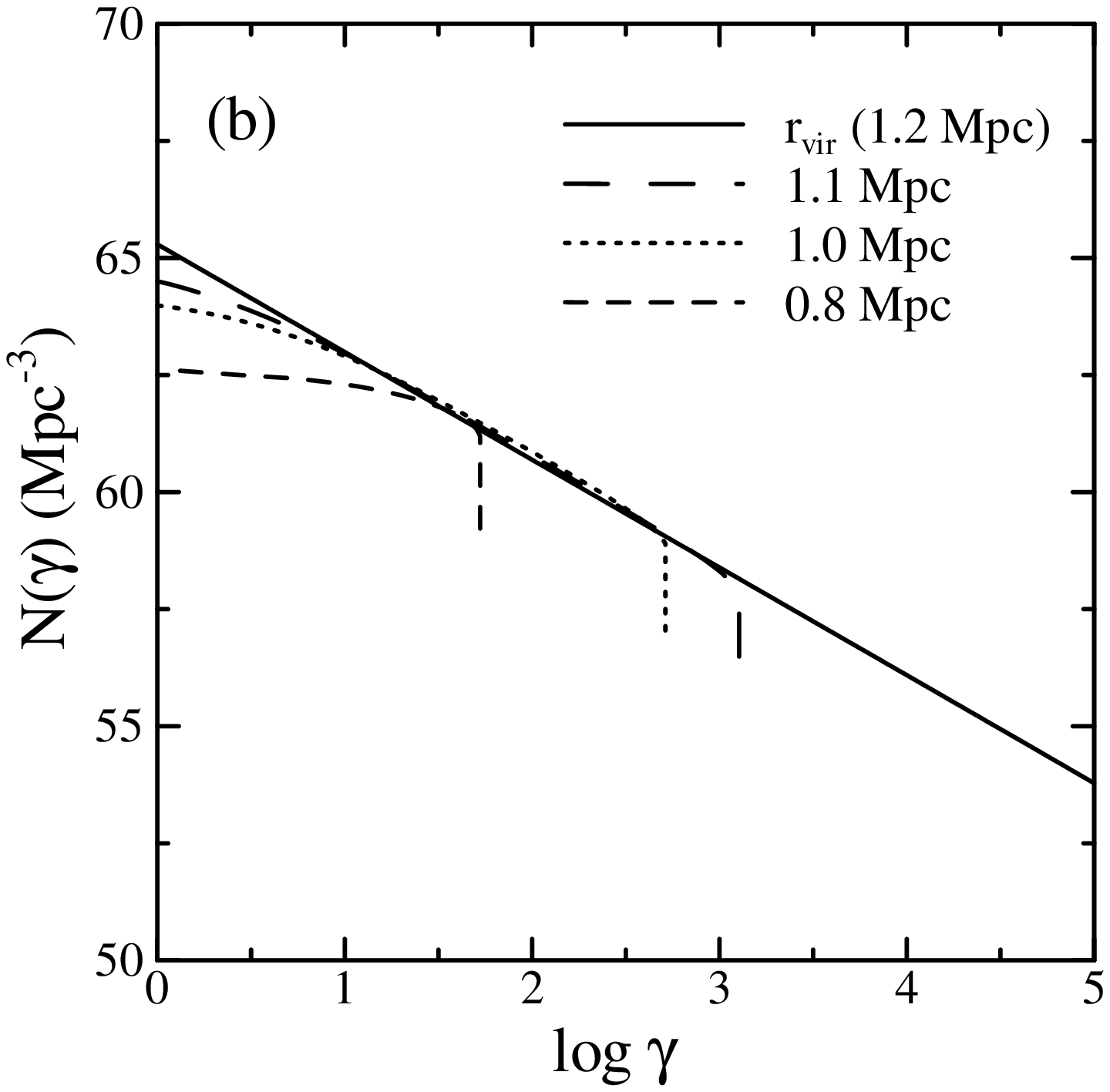} \caption{Radial dependence of electron
spectra. (a) Model~L1 (b) Model~L2. \label{fig:ng}}
\end{figure}

\begin{figure}\epsscale{1.00}
\plottwo{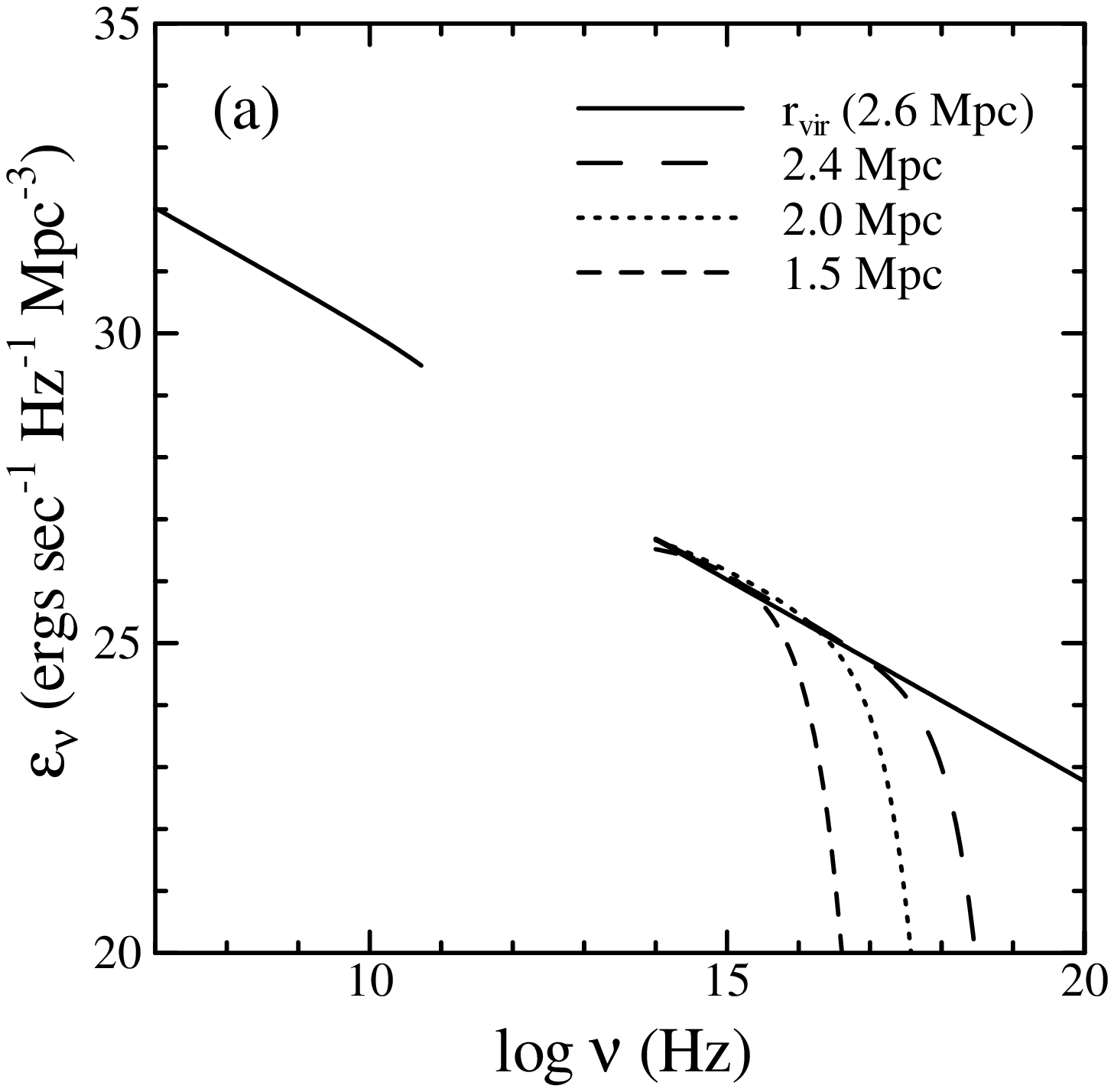}{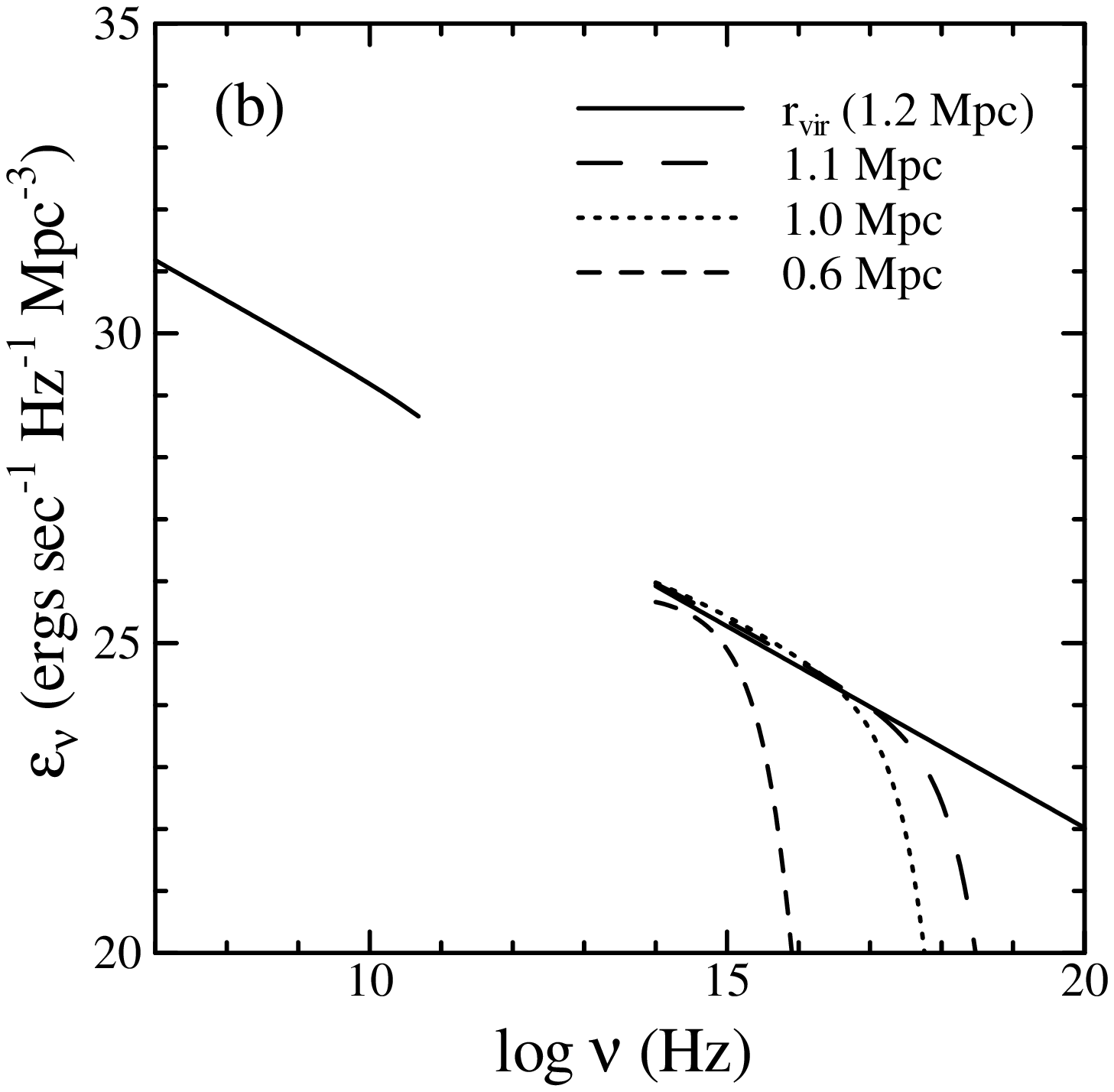} \caption{The radial distributions of the
emissivity as a function of frequency for (a) model~L1 and (b) model~L2.
The emission at $\nu<10^{11}$~Hz is synchrotron emission and that at
$\nu>10^{14}$~Hz is IC emission.  The radio emissivity at all radii
below $r_{\rm vir}$ is too small to be plotted.  \label{fig:rsp}}
\end{figure}

In Figure~\ref{fig:ng}, we present the radial dependence of the electron
spectra. As the radius decreases, electron cooling affects the electron
population because the electrons were accelerated at earlier times. In
particular, IC scattering greatly reduces the number of higher energy
electrons \citep[see][]{sar99}. As a result, the upper cutoff of
electron spectra decreases (Figure~\ref{fig:ng}).  Figure~\ref{fig:rsp}
shows the radial dependence of the synchrotron and IC emissivity of the
cluster as a function of frequency.  In the inner region, the lack of
high-energy electrons reduces the synchrotron and IC emission at higher
frequencies.  In particular, radio synchrotron emission should exist
only at $r\sim r_{\rm vir}$.

\begin{figure}\epsscale{1.00}
\plottwo{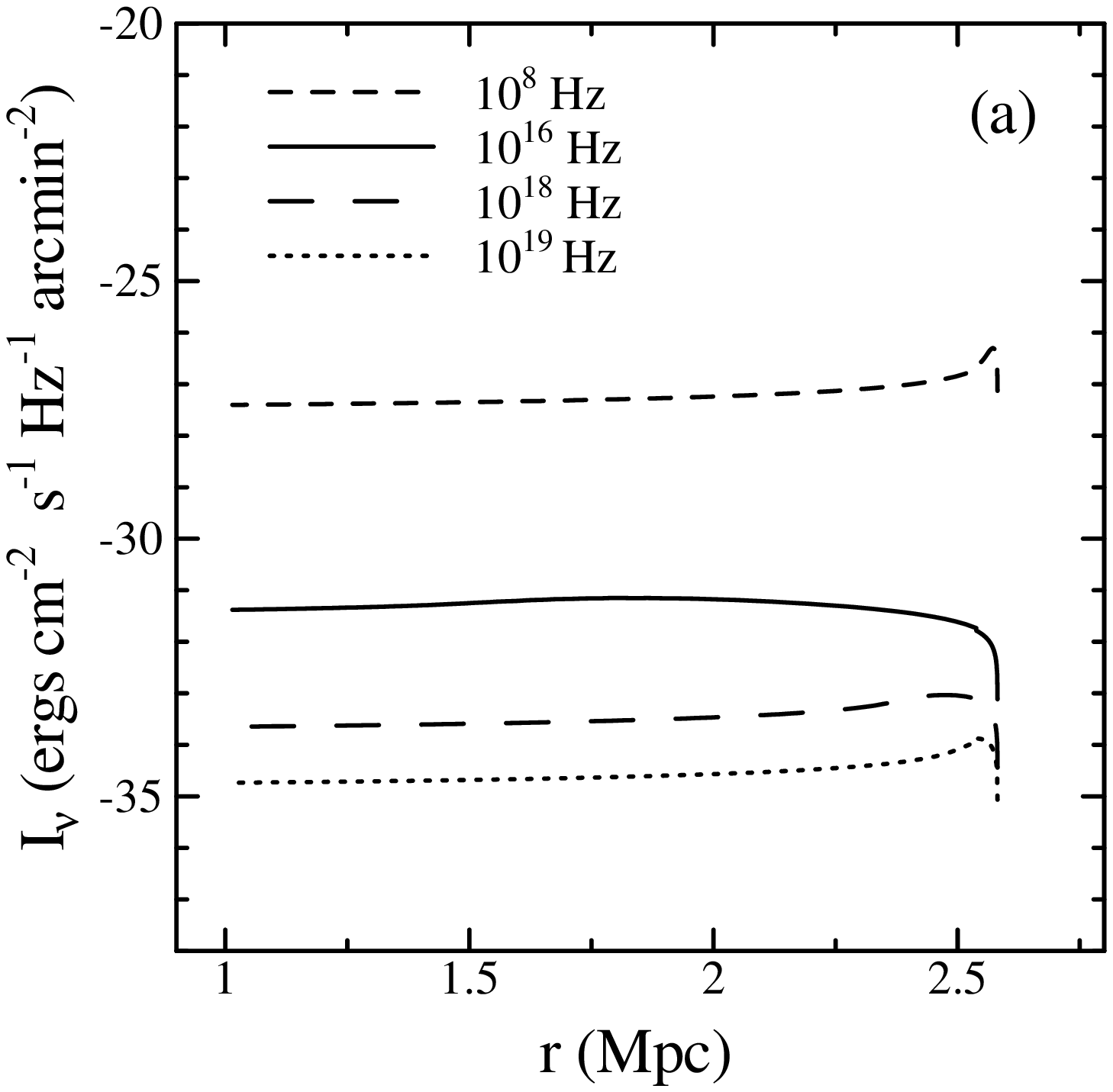}{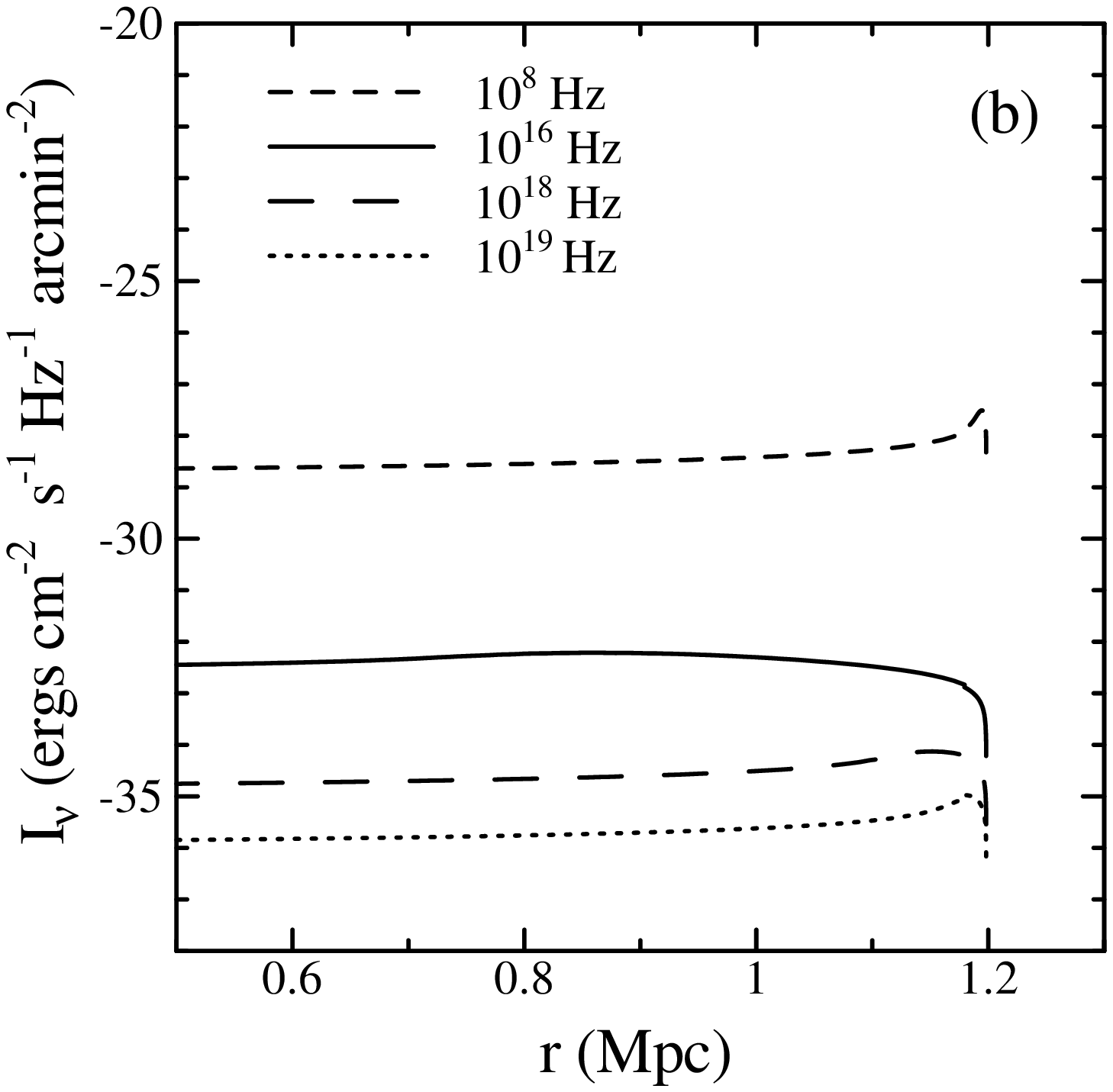} \caption{Surface brightness profiles of
nonthermal emissions for several frequencies. The emission at
$\nu<10^{11}$~Hz is synchrotron emission and that at $\nu>10^{14}$~Hz is
IC emission. (a) Model~L1 (b) Model~L2.\label{fig:Iv}}
\end{figure}

Figure~\ref{fig:Iv} shows the surface brightness profiles of the
nonthermal emission for several frequencies. Since most of the emission,
especially in radio and hard X-ray ranges, comes from the outer region
of clusters (Figure~\ref{fig:rsp}), the profiles are almost constant in
the inner regions. Since synchrotron emission and IC emission at larger
frequency are strictly limited to the cluster edge
(Figure~\ref{fig:rsp}), their surface brightness distributions show
humps near the right ends of the curves or at $r\sim r_{\rm vir}$
(Figure~\ref{fig:Iv}). Figure~\ref{fig:rsp} suggests that the emission
 from the region of $r\lesssim r_{\rm vir}(t_f)$ ($\sim 1.5$~Mpc for
model~L1 and $\sim 0.5$~Mpc for model~L2) does not contribute to the
surface brightness profiles at the frequencies in Figure~\ref{fig:Iv}
except that at $\nu=10^{16}$~Hz in model~L2.

As is mentioned in \S\ref{sec:model}, ``semi-mergers'' ($0.1\lesssim
\Delta_m \lesssim 0.6$) can produce internal shocks. Thus, it is likely
that the nonthermal effects of ``semi-mergers'' penetrate further into
the cluster interior than we calculated on the assumption of
accretion. The effect of semi-mergers on the nonthermal emission due to
the accretion in our sense can be estimated as follows.  Assuming that
the mass evolution of clusters is described by Figure~\ref{fig:m} and
assuming that the average density and magnetic field profiles are given
by Figures~\ref{fig:nr} and~\ref{fig:Br}, respectively, the contribution
of ``pure-accretion'' ($0\lesssim \Delta_m \lesssim 0.1$) to the
nonthermal emission due to the accretion is
\begin{equation}
f_{\rm pure}(t) \approx
\int_{M}^{1.1 M} \Delta M \, r_{\rm LC}^{m}(M \to M'
 ,t) \, dM' \, /
\int_{M}^{1.6 M} \Delta M \, r_{\rm LC}^{m}(M \to M'
 ,t) \, dM' \,,
\end{equation}
where $M=M(t)$ is shown in Figure~\ref{fig:m}. We found that $f_{\rm
pure} = 0.4-0.6$ for essentially all of the models we studied at all
times.  That is, the emission due to the pure accretion is at most
factor of two smaller than that shown Figure~\ref{fig:ts},
\ref{fig:rsp}, and \ref{fig:Iv}. Thus, taking account of the simplicity
of the model, even if we do not include semi-mergers in accretion, the
results of this subsection do not change very significantly.

\subsection{The Emission From a Merging or Merged Cluster}
\label{sec:merger}

\begin{figure}\epsscale{1.00}
\plottwo{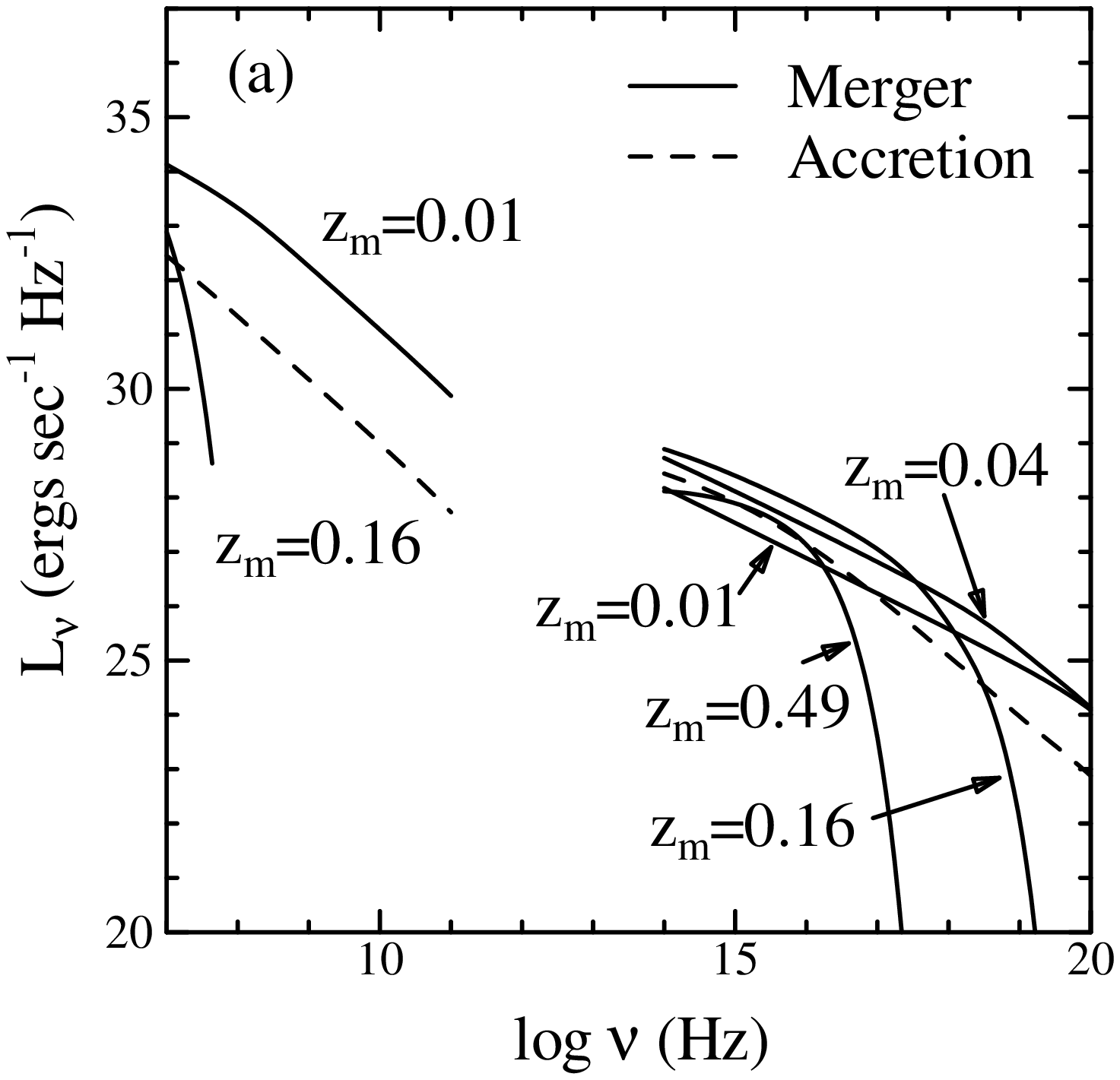}{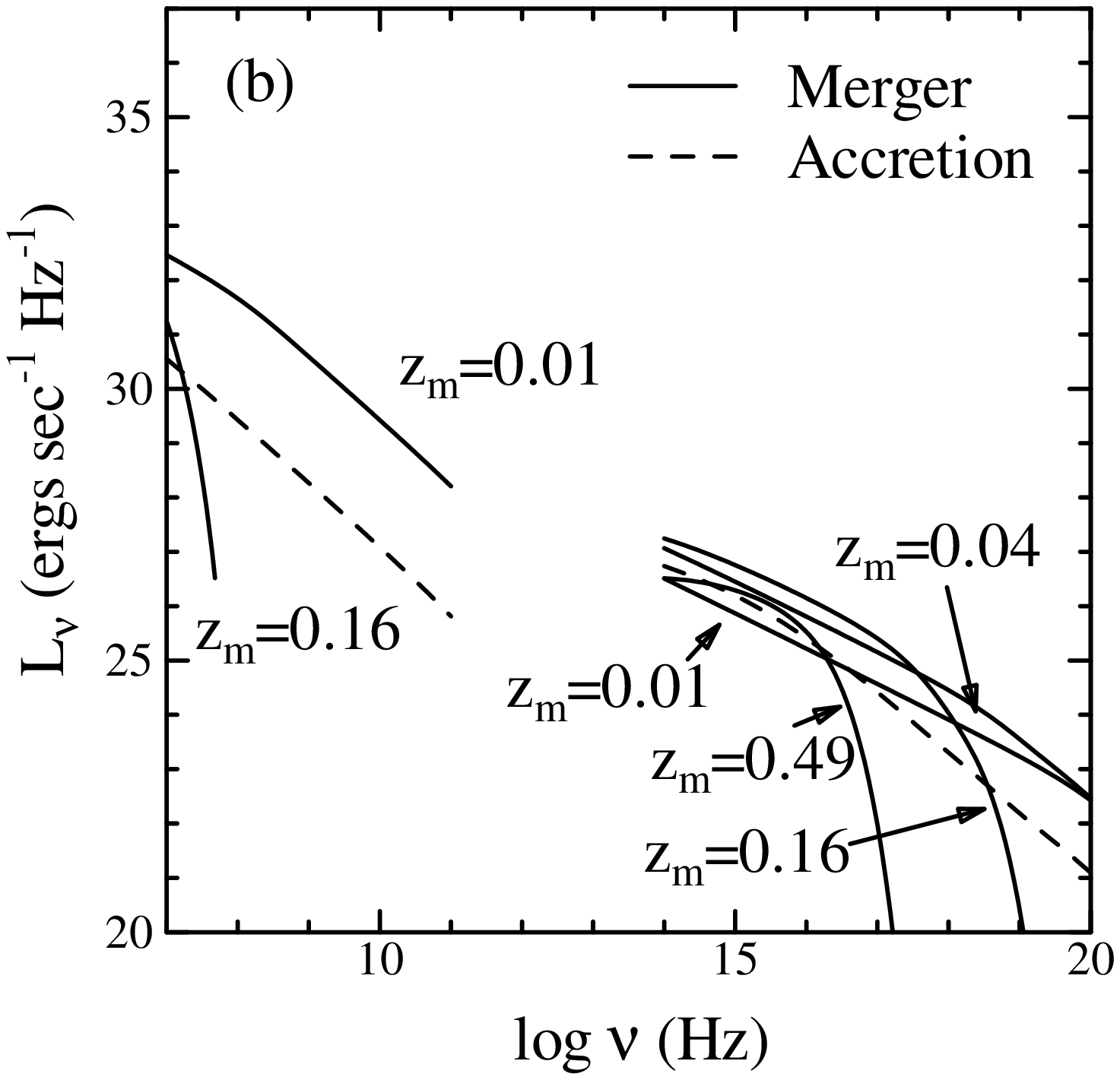} \caption{Spectra of total emission of a
merging or merged cluster for (a) model~L1 and (b) model~L2. The
emission at $\nu<10^{11}$~Hz is synchrotron emission and the emission at
$\nu>10^{14}$~Hz is IC emission. The redshifts indicated in the figure
are that at which the cluster merger starts ($z_m$). For comparison, the
spectra of total emission from a cluster undergoing accretion is also
shown (Figure~\ref{fig:ts}). \label{fig:ts_mer}}
\end{figure}

We assume that there are no nonthermal electrons in a cluster before the
last merger for the sake of simplicity. Figure~\ref{fig:ts_mer} shows
the emission spectra from a merging or merged cluster observed at $z=0$
for models~L1 and~L2; the results of models~S1 and~S2 are similar to
them. The emission from the particles accelerated via accretion is not
included in the emission from the merging or merged cluster; it is shown
separately in Figure~\ref{fig:ts_mer}.

For models~L1 and~L2, a cluster is still in a merging phase at $z=0$ if
the merger starts at $z\lesssim 0.1$. During the merger, the radio
emission and hard X-ray emission ($\nu\gtrsim 10^{19}$~Hz) are almost
constant. This is because the electrons responsible for these emissions
have large energies ($\gamma\gtrsim 10^4$), and thus short cooling times
\citep[$\lesssim 10^8$~yrs; see Fig.~2 in][]{sar99}. Since the energy
injection rate for the nonthermal electrons is assumed to be constant
(\S\ref{sec:mer}), the nonthermal emissions are also constant. However,
the IC emission with $\nu\lesssim 10^{17}$~Hz increases with the
redshift at which the merger starts as long as the cluster is in the
merger phase at $z=0$, because the emission is attributed to the
electrons with $\gamma \lesssim 10^3$ and their cooling times are fairly
long ($\sim 10^9$~yrs); these electrons accumulate in the cluster
without rapidly losing their energy.

For the IC emission, Figure~\ref{fig:ts_mer} shows that the soft X-ray
and EUV emission from a merging cluster is not much different from that
 from an accreting cluster with a similar mass.  For hard X-rays
($\nu\gtrsim 10^{19}$~Hz), the ratio of the luminosity of a merging
cluster to that of an accreting cluster simply reflects the ratio of
their energy injection rates for nonthermal electrons, because the
cooling time of the high-energy electrons responsible for the emissions
is small in comparison with the time scales of merger and accretion. The
ratio of their energy injection rates is approximately given by
\begin{equation}
\frac{L_{\rm X,mer}}{L_{\rm X,acc}}
\sim \frac{GM^2/(r_{\rm vir}^2 t_{\rm shock})}{GM^2/(r_{\rm vir}^2 t_0)}
=\frac{t_0}{t_{\rm shock}}\sim 10
\end{equation}
On the other hand, for $\nu\lesssim 10^{17}$~Hz, the low-energy
nonthermal electrons accumulated in an accreting cluster make the
emission comparable to that from a merging cluster with similar mass.

In comparison with the IC emission, the difference in the synchrotron
emission power between a merging cluster and an accreting cluster is
larger (Figure~\ref{fig:ts_mer}). This is due to the combined effect of
the large rate of energy injection to electrons and the large magnetic
field in the interior of a merging cluster; the magnetic field in the
merging cluster ($\sim 0.1\rm\; \mu G$) exceeds that at the edge of the
accreting cluster by a factor of three. If the ratio of the magnetic
field strength averaged in a cluster to the magnetic field strength at
the cluster edge is much larger than our model, as suggested by
numerical simulations \citep{dol00}, an accreting cluster and a merging
cluster would be even more disparate in radio power.

If a cluster merger starts at $z\gtrsim 0.1$, the merger has completed
before $z=0$. The present radio emission and hard X-ray emission
($\nu\gtrsim 10^{18}$~Hz) due to the last merger rapidly decline with
the redshift of the last merger because of the short cooling time of
high-energy electrons. As soon as the merger finishes, these emissions
become smaller than those due to ongoing accretion. On the contrary, the
IC emission with $\nu\lesssim 10^{17}$~Hz does not change much after the
last merger because of the long cooling time of the relatively
low-energy electrons; even for the cluster that experiences the last
merger at $z\sim 0.4$, the EUV emission is still comparable to that from
the edge of a cluster undergoing accretion.

In Table~\ref{tab:frac}, we list the fraction ($f_{\rm mer}$) of
clusters in a merging phase and the fraction ($f_{\rm mer,EUV}$) whose
EUV luminosity ($\nu\sim 10^{16-17}$~Hz) attributed to their last merger
is larger than that attributed to matter accretion, both at $z=0$.  The
average redshift of cluster formation or the last merger ($z_f$) and the
equal EUV redshift ($z_{\rm EUV}$) are also presented; here, we define
$z_{\rm EUV}$ such that, if a cluster experiences its last merger at
$z<z_{\rm EUV}$, the EUV luminosity due to the last merger is larger
than that due to the accretion at $z=0$. Table~\ref{tab:frac} shows that
only $\sim 10$\% of clusters at $z\sim 0$ are in a merging phase and
radiate strong radio emission due to the merger from their interior.
The remaining clusters radiate weak radio emission due to accretion from
the outskirts. The hard X-ray radiation of the $\sim 10$\% of merging
clusters is relatively strong and it comes from the interior of the
cluster. The hard X-ray radiation of the accreting clusters is weaker
about a factor of 10 and it is limited to their outer
regions. Table~\ref{tab:frac} also suggests that $20-40$\% of clusters
at $z\sim 0$ should have EUV emission due to the last merger from the
whole clusters and the rest should have that from the outer
regions. However, for EUV emission, the distinction between merger and
accretion may be difficult because the EUV emission from an accreting
cluster is relatively uniform and extends to the inner region of the
cluster (Figure~\ref{fig:Iv}).

The nonthermal emission from semi-mergers ($0.1\lesssim \Delta_m
\lesssim 0.6$) may come from the inner regions of clusters
(\S\ref{sec:model}). The luminosity due to nonthermal emission from a
cluster undergoing a semi-merger is expected to be $\gtrsim 10$\% of
that from a cluster undergoing a merger ($\Delta_m \gtrsim 0.6$),
because the typical ICM density and magnetic fields are expected to be
the same for both clusters and the luminosity is proportional to the
kinetic energy of the merged subclusters. Figure~\ref{fig:ts_sem-mer} is
the same as the Figure~\ref{fig:ts_mer} but the merger luminosity is set
to be 20\% of that in Figure~\ref{fig:ts_mer} and the accretion
luminosity is set to be a half of that in Figure~\ref{fig:ts_mer}. That
is, they respectively represent a typical ``semi-merger'' and ``pure
accretion''. As is seen, the luminosity due to a semi-merger is between
that due to pure accretion (or accretion) and that due to a merger for
hard X-ray and radio emissions (Figure~\ref{fig:ts_mer}
and~\ref{fig:ts_sem-mer}). Table~\ref{tab:frac} also shows the fraction
of clusters undergoing semi-mergers or mergers ($f_{\rm smer}$), which
is calculated from the probability distribution, $\Phi_f(z)$ for
$\Delta_m=0.1$.  As can be seen, $f_{\rm smer}\sim 0.2-0.3$ for a low
density universe (models~L1 and~L2) and $f_{\rm smer}\sim 0.3-0.4$ for
the Einstein de-Sitter universe (models~S1 and~S2). Considering the
short cooling time of high energy electrons, these fractions of clusters
should have hard X-ray and radio emissions due to the semi-mergers or
mergers.

\subsection{Magnetic Fields} 
\label{sec:mag}

The magnetic fields derived from observed hard X-ray and EUV emissions
on the assumption that these are IC scattering of CMB photons are rather
low \citep[$\sim 0.2\:\rm \mu G$;][]{fus99,fus00}. On the other hand,
analysis of rotation measure suggests that the magnetic fields in
clusters should be larger a factor of $10-100$
\citep[e.g.,][]{law82,kim90,gol93,cla01}.  Our default parameter
($B_0=0.01\rm\: \mu G$) gives the magnetic fields compatible with the
former. In this subsection, we investigate the magnetic fields
compatible with the latter.

We ran a model with $B_0=0.1\rm\: \mu G$; the other parameters are the
same as in model~L1.  We refer to this model as model~L1$'$. In this
model, the magnetic field is 10 times as large as that in
model~L1. Thus, for an accreting cluster, the magnetic field in the
central region is $\sim 10\rm\; \mu G$ and that at the cluster edge is
$0.5\rm\; \mu G$. The magnetic field of a merging cluster is $1.3\rm\;
\mu G$. Figure~\ref{fig:ts_mag}a shows the total emission from the
cluster. The IC emission in model~L1$'$ is not much different from that
in model~L1 because the IC emission is not much affected by the change
of magnetic fields as long as $B\lesssim 1\;\mu G$ \citep{sar99}. On the
contrary, the synchrotron luminosity in model~L1$'$ is much larger than
that in model~L1, because the synchrotron emission is proportional to
$B^2$ for a given population of high-energy electrons.

\section{Discussion}
\label{sec:dis}

In this section, we compare the results in \S\ref{sec:res} with
observations. We use the cosmological parameters for model~L
(Table~\ref{tab:par}) or $\Omega_0=0.7$, $\lambda=0.3$ and $h=0.7$.
Unfortunately, excess EUV and hard X-ray fluxes have been detected from
only a few clusters so far, and thus we cannot discuss the emission
statistically. Instead, we check the crude consistency between our
simple model and observations for the individual clusters. On the other
hand, thanks to the better sensitivity of radio telescopes, the number
of clusters in which diffuse radio emission is detected is increasing,
although the numbers may still be a bit small for detailed statistical
analysis.

\subsection{EUV/Soft X-ray Emission}

Extreme ultraviolet (EUV) or very soft X-ray excess emission have been
detected from a number of clusters
\citep{lie96a,lie96b,bow96,bow98,mit98,kaa99,lie99a,lie99b,ber00a,
bon01}. although it should be noted that the EUV detections remain
controversial \citep{ara99,bow99,ber00b}.  \citet{bon01} detected excess
EUV emission from A1795 and found that the total EUV luminosity is $\sim
3 \times 10^{43}$ ergs s$^{-1}$ for $60-250$~eV. The luminosity is
consistent with model~L1 series for both accretion and merger
(Figure~\ref{fig:ts_mer}a, \ref{fig:ts_sem-mer}a
and~\ref{fig:ts_mag}). However, the observed emission is restricted to
the region of $r\lesssim 700$~kpc for the cosmological parameters we
used, which is inconsistent with the spread of the emission that our
accretion model predicts (Figure~\ref{fig:Iv}a). As for A2199, Coma, and
Virgo cluster, the same can be said \citep{lie99b,bow98,ber00b}. For an
example, \citet{kaa99} found excess emission from cluster A2199 in the
$0.1-0.3$~keV band. Although the luminosity, $\sim 6 \times 10^{42}$
ergs $^{-1}$, is consistent with model~L1 series
(Figure~\ref{fig:ts_mer}a, \ref{fig:ts_sem-mer}a, and~\ref{fig:ts_mag}),
the emission is mainly limited to the region of $\lesssim 400$~kpc and
is inconsistent with our prediction in the case of accretion
(Figure~\ref{fig:Iv}a). These results suggest that the observed EUV
emission is at least not attributable to the outer accretion shocks at
the virial radius in clusters. On the other hand, the surface brightness
of the EUV/soft X-ray emission produced by accretion is quite low, and
is probably below the detection limit of the existing observations.
Thus, the observations probably do not rule out accretion generated
emission at the level predicted by the models, and even the reports of
non-detection of EUV in clusters \citep{ara99,bow99,ber00b} may not be
inconsistent with the accretion models.

\subsection{Hard X-ray Emission}
\label{sec:hard}

\begin{figure}\epsscale{1.00}
\plottwo{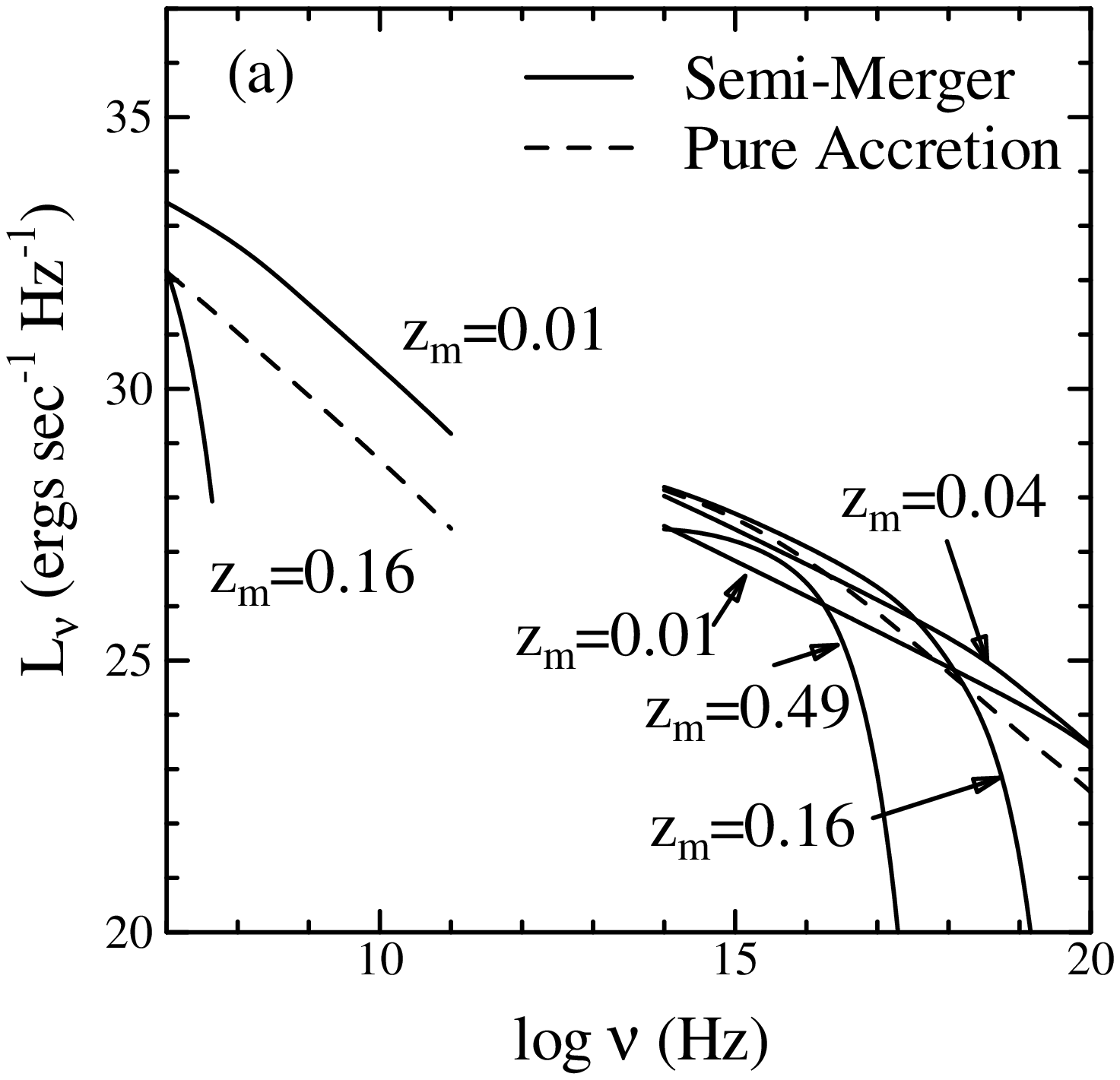}{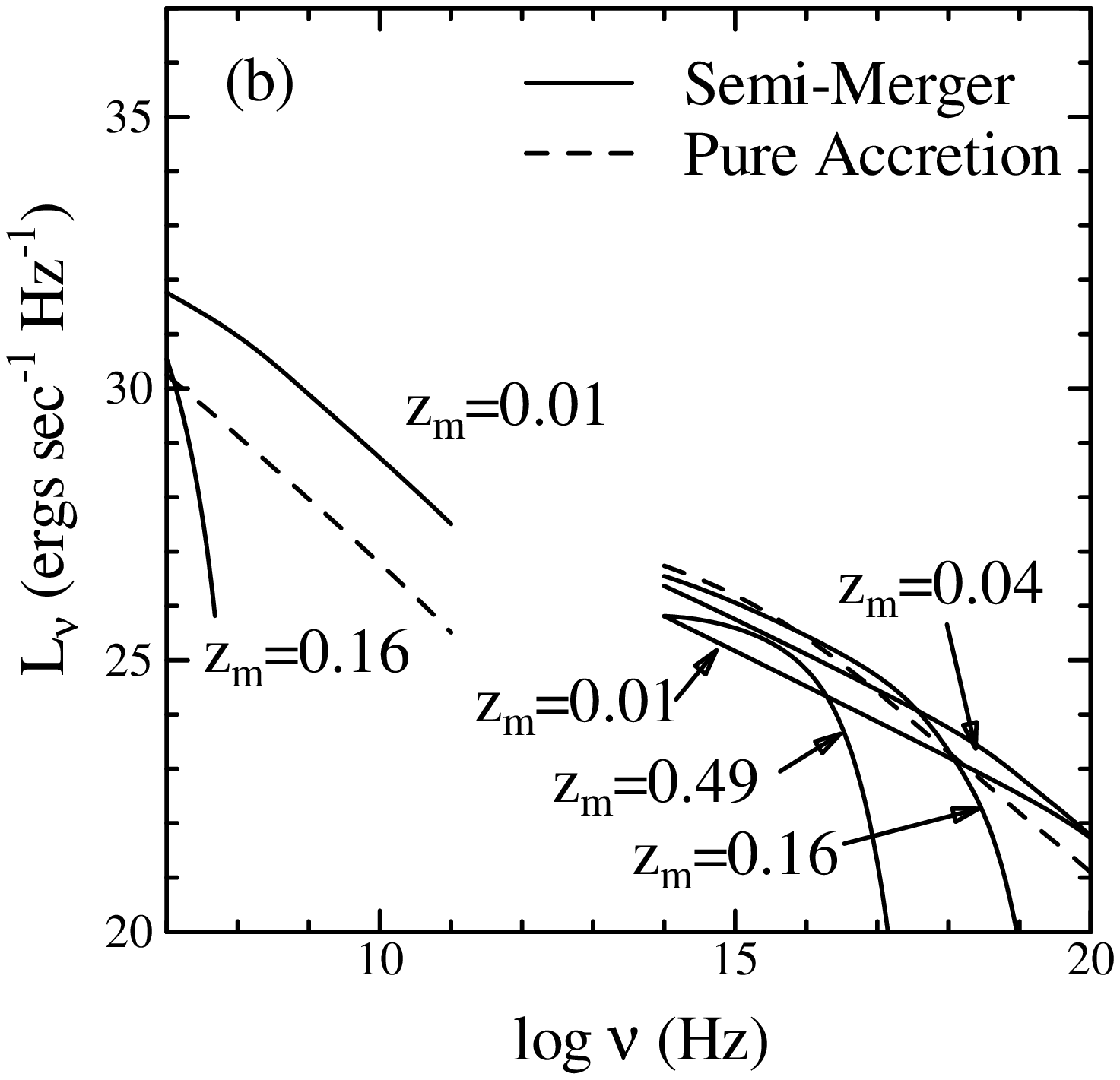} \caption{The same as
Figure~\ref{fig:ts_mer} but for semi-mergers and pure accretion. (a)
model~L1 and (b) model~L2. \label{fig:ts_sem-mer}}
\end{figure}

\begin{figure}\epsscale{0.50}
\plotone{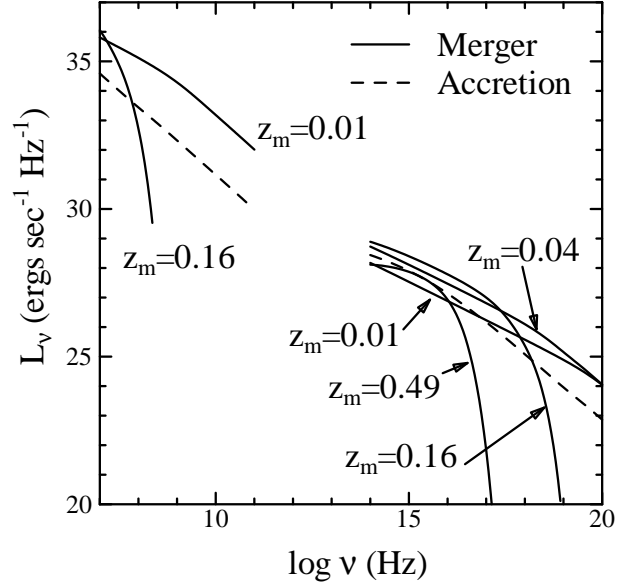} \caption{The same as Figure~\ref{fig:ts_mer}a except for
$B_0=0.1\rm\: \mu G$ (model~L1$'$). \label{fig:ts_mag}}
\end{figure}

Excess hard X-ray emission has been detected in two clusters, Coma and
Abell~2256.  There is weaker evidence for hard X-ray excesses in
Abell~2199 \citep{kaa99} and Abell~3667 \citep{fus01}.  In Coma, the
detection was made with both {\it BeppoSAX} \citep{fus99} and {\it RXTE}
\citep{rep99}.  For Coma cluster, the {\it BeppoSAX} hard X-ray flux is
$2.2\times 10^{-11}$ ergs cm$^{-2}$ s$^{-1}$ in the $20-80$~keV band.
This corresponds to the luminosity of $\sim 10^{43}$ ergs s$^{-1}$,
which is consistent with the prediction of model~L1 for accretion
(Figure~\ref{fig:ts_mer}a, \ref{fig:ts_sem-mer}a
and~\ref{fig:ts_mag}). However, it is smaller than the model for a major
merger ($\sim 10^{44}$ ergs s$^{-1}$), although Coma Cluster is known as
a merging cluster \citep[e.g.,][]{bri92,vik97,hon96,wat99}. The small
luminosity suggests that the ratio of mass between the main cluster and
the merger subcluster or subclusters fairly be large.  For Abell~2256,
which is also known as a merging cluster
\citep[e.g.,][]{bri91,miy93,bri94,mar96,mol00}, the flux of hard X-ray
emission is $1.2 \times 10^{-11}$ ergs cm$^{-2}$ s$^{-1}$ in the
$20-80$~keV band range \citep{fus00}. The corresponding luminosity is
$\sim 10^{44}$ ergs s$^{-1}$, which is consistent with the prediction of
model~L1 series for a merger (Figure~\ref{fig:ts_mer}a,
\ref{fig:ts_sem-mer}a and~\ref{fig:ts_mag}). The observed radio
luminosities for Coma and A2256 are $1 \times 10^{40}$ and $2 \times
10^{40}$ erg s$^{-1}$ at 1.4~GHz, respectively \citep{fer00}. Both these
clusters have both a radio halo and a relic.  By comparing the
luminosities with the model~L1 series, it can be shown that $B_0$ must
be $0.01\:\rm \mu G$ or the average magnetic fields in the clusters much
be $\bar{B}\sim 0.1\:\rm \mu G$ whether the emission is due to a merger
or to accretion (Figure~\ref{fig:ts_mer}a, \ref{fig:ts_sem-mer}a
and~\ref{fig:ts_mag}). Future observations of the spatial distributions
of the hard X-ray emission would be quite useful to investigate whether
mergers or accretion are responsible for the hard X-ray emission, and
whether the hard X-ray emission is spatially coincident with the radio
halos and/or the relics.

\citet{fuk01} detected an excess of hard X-ray at energies above 4~keV
 from the group of galaxies HCG~62. In the $2-10$~keV range, the observed
hard X-ray flux is $(1.0\pm 0.3)\times 10^{-12}\rm\: ergs\: cm^{-2}\:
s^{-1}$. This implies a luminosity of $\sim 4\times 10^{41}\rm\: ergs\:
s^{-1}$, which is consistent with the accretion of model~L2
(Figure~\ref{fig:ts_mer}b and~\ref{fig:ts_sem-mer}b). However, the
emission is confined to the region of $r\lesssim 200$~kpc, which is
inconsistent with the result in Figure~\ref{fig:Iv}b (model~L2). Thus,
we can say at least that the observed hard X-ray emission is not
produced by electron acceleration through matter accretion to the
group. Note that there is no strong evidence of merger for this group
\citep{fuk01}.  There is no radio detection of a diffuse source in this
group.

\subsection{Radio Emission}

In some clusters, radio relics and halos are observed. Considering their
locations, the relics may be the synchrotron emission from accelerated
electrons at the cluster edge. On the other hand, the radio halos may be
produced by mergers. \citet{ens98b} compiled the observational results
of relics and showed that the luminosities are $\sim 10^{40-42}\rm\:
ergs\: s^{-1}$ assuming a simple power law between 10~MHz and
10~GHz. Recently, \citet{gio99} searched for new halo and relic
candidates in the NRAO VLA Sky Survey and found 29 candidates; the
luminosities of the candidates are $\sim 10^{39-41}\rm\: ergs\: s^{-1}$
at $\nu=1.4$~GHz. Moreover, \citet{kem01} also searched for halos and
relics in all of the Abell clusters that are visible in the Westerbork
Northern Sky Survey and found 18 candidates; the luminosities of the
candidates are $\sim 10^{39.5-40.5}\rm\: ergs\: s^{-1}$ at
$\nu=327$~MHz. Furthermore, \citet{gio00} also find diffuse radio
emissions from 7 clusters with VLA and the luminosities are $\sim
10^{39.5-40.5}\rm\: ergs\: s^{-1}$ at $\nu=0.3$ and 1.4~GHz. These are
consistent with model~L1 (Figure~\ref{fig:ts_mer}a and
Figure~\ref{fig:ts_sem-mer}a); the dispersion of luminosities can be
explained by the difference between merger and accretion. On the other
hand, the observed luminosities are not consistent with model~L1$'$
(Figure~\ref{fig:ts_mag}) as the clusters in which hard X-ray emission
is detected. These may suggest that magnetic fields in clusters are {\rm
generally} small ($B_0\sim 0.01\rm\: \mu G$ or $\bar{B}\sim 0.1\:\rm \mu
G$).

\begin{figure}\epsscale{0.50}
\plotone{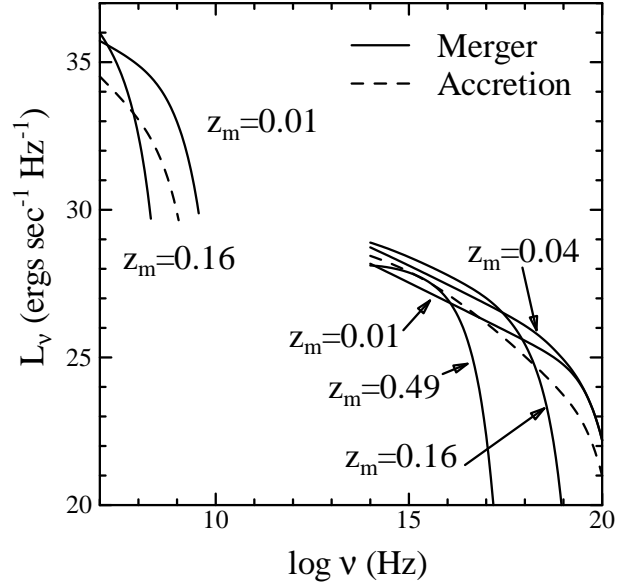} \caption{The same as Figure~\ref{fig:ts_mag} except
that the initial electron energy distribution has a high-energy cutoff for
$\gamma_i > 10^4$. \label{fig:cut}}
\end{figure}

We note that such weak magnetic fields are not consistent with the large
values implied by Faraday rotation measurements
\citep[e.g.,][]{law82,kim90,gol93,cla01}.  Our results suggest that this
discrepancy exists generally for clusters. Several authors have
suggested ways to reconcile this difference, such as spatial
inhomogeneity of the magnetic field and/or the relativistic particles
\citep{gol93}, nonthermal bremsstrahlung emission by semi-relativistic
particles as the source of the hard X-ray emission in clusters rather
than IC \citep{ens99,sar00}, an anisotropic pitch angle distribution of
electrons \citep{pet01}, and a high-energy cutoff in the electron energy
distribution \citep{bru01,pet01}.  It is beyond the scope of this paper
to consider all of these possibilities in concert with the accretion and
merger models we give. However, as an example we consider electron
energy distributions with a high-energy cutoff.  We construct a simple
model with the same parameters as model~L1$'$, but with $N_i=0$ for
$\gamma_i>10^4$; such an electron distribution may be achieved when
electrons accelerated in the past are reaccelerated by mergers at
present \citep{bru01}. The result is shown in
Figure~\ref{fig:cut}. Compared to Figure~\ref{fig:ts_mag}, radio
luminosity is much smaller, while the IC emission is not much different
for $\nu<10^{19}$~Hz. Thus, a higher magnetic field could be
accommodated by this model.

Most of the observed relics are not spherically symmetric.  Thus, some
of them may not have originated from pure accretion but from mergers or
semi-mergers.  The latter is consistent with the fact that radio relics
are not observed in all clusters but only those with other evidence for
mergers, although the incompleteness of the existing samples is a
concern.  The lack of symmetry might indicate that both an accretion
shock and a preexisting relativistic population (a ``radio ghost'') may
be required to produce a relic \citep{ens01}.  On the other hand, given
the limited sensitivity and incompleteness of radio searches for relics,
it is possible that most clusters have low surface brightness, nearly
spherically symmetric relics due to accretion shocks.  Also, nonthermal
emission from accretion shocks may not be generally spherically
symmetric.  This might occur because the accretion is always associated
with clustered matter, and thus occurs through the intermittent
accretion of small blobs ($\Delta_m \lesssim 0.1$).  Given the short
cooling time of high-energy electrons, this would lead to asymmetric
radio emission (Figures~\ref{fig:ng} and \ref{fig:rsp}).  Alternatively,
accretion may be asymmetric if it occurs largely from filamentary large
scale structures \citep{ens98b}.  In this case, the sense of the
asymmetries would tend to remain constant over time, and would align
with the large scale structure.  If the radio emission due to accretion
is asymmetric, one would expect a similar asymmetric structure in the
hard X-ray IC emission as well, because this emission originates from
the same higher energy electrons as the synchrotron emission.

\section{Conclusions}
\label{sec:conc}

According to the theories of structure formation in the universe, most
clusters at $z\sim 0$ are not in a merging phase but are undergoing
accretion.  Thus, we have investigated the nonthermal emission from the
electrons accelerated by shocks around the cluster edge formed through
matter accretion to the cluster. We compared the emission with the
nonthermal emission from merging clusters. Moreover, we have shown the
radial profiles of the nonthermal emission for accreting clusters. In
order to estimate the accretion rate and the probability distribution of
cluster merger, we used a modification of an extend Press \& Schechter
model. We assumed that the rate of energy injection to high-energy
electrons is proportional to the released gravitational energy of
accreted matter or subclusters. We considered the energy loss through
inverse Compton (IC) scattering of cosmic microwave background (CMB),
synchrotron radiation, Coulomb losses, and bremsstrahlung radiation.

The results of our calculations show that the nonthermal emission owing
to the accretion is restricted to the outer region of clusters. In
particular, the synchrotron emission and the IC emission at high
frequency ($\nu\gtrsim 10^{19}$~Hz) are limited to the periphery of
clusters. Thus, the surface brightness profiles of the nonthermal
emissions are almost flat in the inner region of clusters regardless of
frequency. On the other hand, the surface brightness of the synchrotron
emission and the IC emission at high frequency ($\nu\gtrsim 10^{19}$~Hz)
is enhanced at the cluster edge. The total IC emission from an accreting
cluster is not much different from that from a merging cluster; for EUV
emission, they are comparable, and for hard X-ray emission, the former
is smaller than the latter a factor of 10. On the other hand, the
synchrotron emission from an accreting cluster is about 100 times
smaller than that from a merging cluster. We predict that about 10\% of
clusters at $z\sim 0$ have hard X-ray and radio nonthermal emissions due
to their last merger, which are comparable to or dominate those due to
ongoing accretion. 
On the other hand, if we define a merger as an increase in the mass of
the larger subcluster by at least 10\% of its initial mass, about
$20-30$\% ($30-40$\%) of clusters at $z\sim 0$ should have hard X-ray
and radio nonthermal emissions due to the merger even in a low density
universe (in the Einstein de-Sitter universe). We also predict that
$20-40$\% of clusters have significant extreme ultraviolet (EUV)
emission due to their last major merger.

We compare the results with observations of clusters. We find that both
our accretion and merger models are energetically consistent with the
EUV emission of clusters. However, the observed emission is concentrated
to the cluster centers in contrast with our prediction of accretion. The
total luminosity of hard X-ray emission via IC scattering appears to be
consistent with the observations of clusters. By comparing the observed
hard X-ray emission and radio synchrotron emission for individual
clusters, it is shown that the average magnetic fields should be $\sim
0.1\rm\: \mu G$ for our simple model.  Recently, the number of clusters
in which diffuse radio emission is detected is increasing. If the
emission is synchrotron emission due to accretion or mergers, the
magnetic energy of clusters may generally be as small as in the clusters
in which hard X-ray emission has been detected. However, we also find
that models with a high-energy cutoff in the electron energy
distribution will accommodate larger magnetic fields, which are in
better agreement with Faraday rotation measurements.

\acknowledgments 

We thank T. Furusyo, N. Y. Yamasaki, T. Ohashi, M. Yamada, and M. Enoki
for useful comments.  C. L. S. was supported in part by by the National
Aeronautics and Space Administration through {\it XMM} grant NAG5-10075
and {\it Chandra} Award Numbers through {\it Chandra} Award Number
GO0-1158X, GO0-1173X, GO1-2122X, and GO1-2123X.  all issued by the {\it
Chandra} X-ray Observatory Center, which is operated by the Smithsonian
Astrophysical Observatory for and on behalf of NASA under contract
NAS8-39073.

\clearpage


\clearpage

\clearpage

\begin{deluxetable}{lcccccc}
\tablecaption{Model Parameters. \label{tab:par}}
\tablewidth{0pt}
\tablehead{
\colhead{Models} & \colhead{$\Omega_0$}   & \colhead{$\lambda$}   &
\colhead{$h$} &
\colhead{$\sigma_8$}  & \colhead{$M_0$} & \colhead{$B_0$} 
}
\startdata
L1   & 0.3 & 0.7 & 0.7 & 1.00 & $10^{15}\:M_{\sun}$ & $0.01\rm\: \mu G$\\
L1$'$& 0.3 & 0.7 & 0.7 & 1.00 & $10^{15}\:M_{\sun}$ & $0.10\rm\: \mu G$\\
L2   & 0.3 & 0.7 & 0.7 & 1.00 & $10^{14}\:M_{\sun}$ & $0.01\rm\: \mu G$\\
S1   & 1.0 & 0.0 & 0.5 & 0.63 & $10^{15}\:M_{\sun}$ & $0.01\rm\: \mu G$\\
S2   & 1.0 & 0.0 & 0.5 & 0.63 & $10^{14}\:M_{\sun}$ & $0.01\rm\: \mu G$\\
\enddata

\end{deluxetable}


\begin{deluxetable}{cccccc}
\tablecaption{Merger Fractions in Clusters \label{tab:frac}}
\tablewidth{0pt}
\tablehead{
\colhead{Models} & $f_{\rm mer}\tablenotemark{a}$ & 
$f_{\rm mer,EUV}\tablenotemark{b}$   &
\colhead{$z_f$\tablenotemark{c}} &
\colhead{$z_{\rm EUV}$\tablenotemark{d}} &
\colhead{$f_{\rm smer}\tablenotemark{e}$}
}
\startdata
L1 & 0.09 & 0.3 & 0.66 & 0.4 & 0.31\\
L2 & 0.05 & 0.2 & 1.10 & 0.4 & 0.21\\
S1 & 0.14 & 0.4 & 0.36 & 0.3 & 0.44\\
S2 & 0.07 & 0.2 & 0.77 & 0.3 & 0.30\\
\enddata

\tablenotetext{a}{Fraction of clusters in a merging phase at $z=0$.}
\tablenotetext{b}{Fraction of the clusters whose EUV luminosities
attributed to their last merger is larger than those attributed to
accretion at $z=0$.} \tablenotetext{c}{Average redshift of cluster
formation.} \tablenotetext{d}{Equal EUV redshift (see
text)}.\tablenotetext{e}{Fraction of clusters in a semi-merging or
merging phase at $z=0$.}

\end{deluxetable}

\end{document}